%% file: 00_abstract.tex
\documentclass[manuscript,screen]{acmart}

\usepackage{hyperref}
\usepackage{amsmath}
\usepackage{graphicx}
\usepackage{textcomp}
\usepackage{xcolor}
\usepackage{listings}
\usepackage[noend]{algpseudocode}
\usepackage{multirow}
\usepackage{booktabs}
\usepackage{lipsum}
\usepackage{threeparttable}
\usepackage{hhline}
\usepackage{array}
\usepackage{subcaption}

\usepackage{xspace}
\usepackage{booktabs}
\usepackage{threeparttable}
\usepackage{tablefootnote}

\AtBeginDocument{%
  }

\definecolor{LightBlue}{rgb}{0.83, 0.91, 1}
\definecolor{Red}{rgb}{1, 0, 0}

\newcommand{\SupportCore}{\emph{support-core}\xspace}

\newcommand{\ASNGMalloc}{\textit{SpeedMalloc}\xspace}

\newcommand{\JeSpeedup}{$1.75 \times$}
\newcommand{\TCSpeedup}{$1.18 \times$}
\newcommand{\MiSpeedup}{$1.15 \times$}
\newcommand{\MaSpeedup}{$1.23 \times$}
\newcommand{\MeSpeedup}{$1.18 \times$}

\newcommand{\JeSpeedupP}{$42.93\%$}
\newcommand{\TCSpeedupP}{$15.30\%$}
\newcommand{\MiSpeedupP}{$13.06\%$}
\newcommand{\MaSpeedupP}{$18.71\%$}
\newcommand{\MeSpeedupP}{$15.02\%$}

\newcommand{\Allocation}{$4.4\%$}

\newcommand{\JeEnergy}{$1.69 \times$}
\newcommand{\TCEnergy}{$1.15 \times$}
\newcommand{\MiEnergy}{$1.12 \times$}
\newcommand{\MaEnergy}{$1.26 \times$}
\newcommand{\MeEnergy}{$1.22 \times$}

\begin{document}

\setcopyright{none}
\settopmatter{printacmref=false} 
\renewcommand\footnotetextcopyrightpermission[1]{} 

\title{SpeedMalloc: Improving Multi-threaded Applications via a Lightweight Core for Memory Allocation
}

\author{Ruihao Li}
\email{liruihao@utexas.edu}
\affiliation{%
  \institution{The University of Texas at Austin}
  \city{Austin}
  \state{Texas}
  \country{USA}
}

\author{Qinzhe Wu}
\email{qw2699@utexas.edu}
\affiliation{%
  \institution{University of Texas at Austin}
  \city{Austin}
  \state{Texas}
  \country{USA}
}

\author{Krishna Kavi}
\email{Krishna.Kavi@unt.edu}
\affiliation{%
  \institution{University of North Texas}
  \city{Denton}
  \state{Texas}
  \country{USA}
}

\author{Gayatri Mehta}
\email{Gayatri.Mehta@unt.edu}
\affiliation{%
  \institution{University of North Texas}
  \city{Denton}
  \state{Texas}
  \country{USA}
}

\author{Jonathan C. Beard}
\email{jonathan.c.beard@gmail.com}
\authornote{Currently in Google.}
\affiliation{%
  \institution{ARM}
  \city{Austin}
  \state{Texas}
  \country{USA}
}

\author{Neeraja J. Yadwadkar}
\email{neeraja@austin.utexas.edu }
\affiliation{%
  \institution{The University of Texas at Austin}
  \city{Austin}
  \state{Texas}
  \country{USA}
}

\author{Lizy K. John}
\email{ljohn@ece.utexas.edu}
\affiliation{%
  \institution{The University of Texas at Austin}
  \city{Austin}
  \state{Texas}
  \country{USA}
}

\renewcommand{\shortauthors}{Ruihao Li et al.}

\begin{abstract}
Memory allocation, though constituting only a small portion of the executed code, can have a "butterfly effect" on overall program performance, leading to significant and far-reaching impacts.
Despite accounting for just $\sim$$5\%$ of total instructions, memory allocation can result in up to a $2.7\times$ performance variation depending on the allocator used.
This effect arises from the complexity of memory allocation in modern multi-threaded multi-core systems, where allocator metadata becomes intertwined with user data, leading to cache pollution or increased cross-thread synchronization overhead.
Offloading memory allocators to accelerators, e.g., \textit{Mallacc} and \textit{Memento}, is a potential direction to improve the allocator performance and mitigate cache pollution.
However, these accelerators currently have limited support for multi-threaded applications, and synchronization between cores and accelerators remains a significant challenge. 
\par

We present \ASNGMalloc, using a lightweight \SupportCore to process memory allocation tasks in multi-threaded applications.
The \SupportCore is a lightweight programmable processor with efficient cross-core data synchronization and houses all allocator metadata in its own caches.
This design minimizes cache conflicts with user data and eliminates the need for cross-core metadata synchronization.
In addition, using a general-purpose core instead of domain-specific accelerators makes \ASNGMalloc capable of adopting new allocator designs.
We compare \ASNGMalloc with state-of-the-art software and hardware allocators, including Jemalloc, TCMalloc, Mimalloc, \textit{Mallacc}, and \textit{Memento}. 
\ASNGMalloc achieves \JeSpeedup, \TCSpeedup, \MiSpeedup, \MaSpeedup, and \MeSpeedup \space speedups on multithreaded workloads over these five allocators, respectively.
\end{abstract}

\maketitle

\input{content}

\bibliographystyle{ACM-Reference-Format}
\bibliography{refs}

\end{document}

%% file: content.tex
\section{Introduction}
\label{section_introduction}
\input{01_introduction}

\section{ Related Work \& Motivation}
\label{section_background}
\input{02_background}

\section{\ASNGMalloc Overview}
\label{section_overview}
\input{03_overview}

\section{\ASNGMalloc Invocation Interface}
\label{section_control}
\input{04_control}

\section{\ASNGMalloc Runtime}
\label{section_data_transfer}
\input{05_data}

\section{Evaluation}
\label{section_evaluation}
\input{08_evaluation}

\section{Conclusion}
\label{section_conclusion}
\input{010_conclusion}

%% file: 01_introduction.tex
\begin{figure}[t]
\centerline{\includegraphics[width=0.7\columnwidth,trim = 15mm 8mm 19mm 8mm, clip=true, page=1]{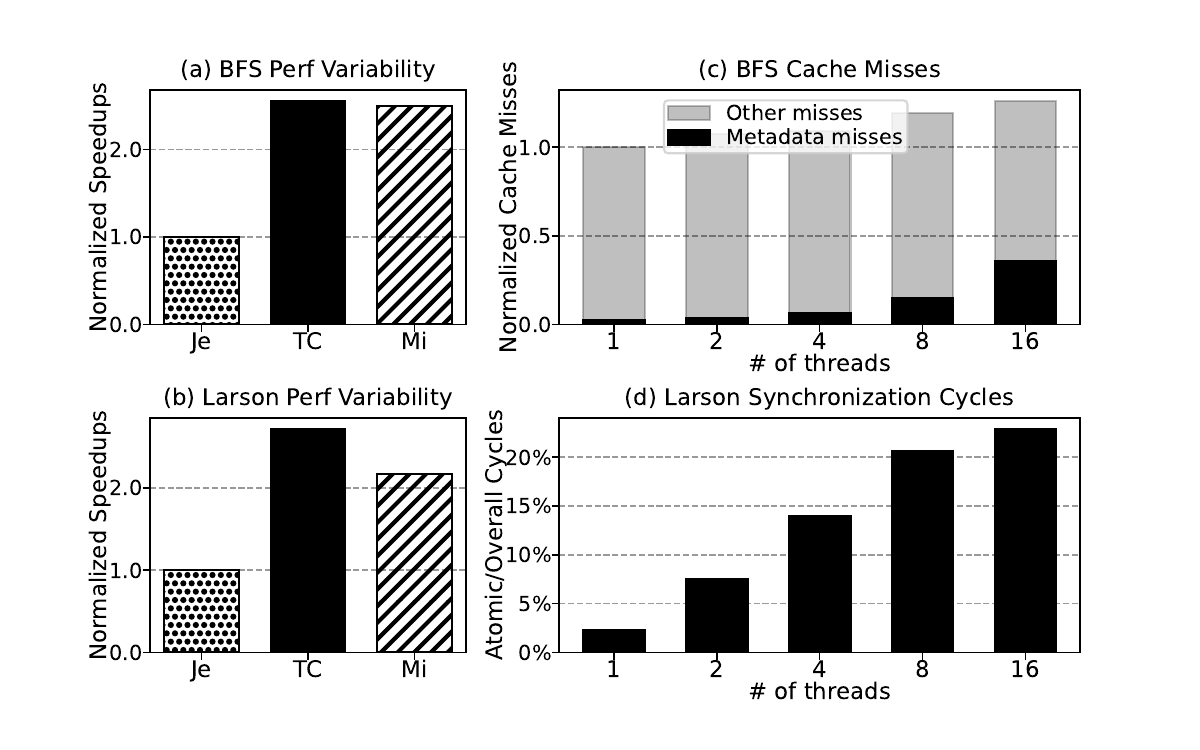}}
	\caption{
    Performance varies significantly depending on the allocator. Memory allocators have a significant impact on the total execution time as it affects metadata cache and cross-core synchronization behaviors.
    (a) $2.55 \times$ performance variability between allocators for \textit{BFS}.
    (b) $2.71 \times$ performance variability between allocators for \textit{Larson}.
    (c) Profiling TCMalloc for \textit{BFS} shows that conflicts with allocator metadata lead to $28.3\%$ of all cache misses when using 16 threads.
    (d) Profiling TCMalloc for \textit{Larson} shows that 22.8\% of execution cycles are spent on allocator metadata synchronization (16-thread run). 
    }
	\label{fig_bfs}
\end{figure}

Memory management and allocation consume a substantial proportion of datacenter computation cycles~\cite{kanev2015profiling, sriraman2020accelerometer, gonzalez2023profiling, zhou2024characterizing}. 
Prior studies found that allocators impact the performance not only of allocation tasks but also the entire program~\cite{li2023nextgen} and even the entire data center fleet~\cite{hunter2021beyond, zhou2024characterizing}. 
Prior works have shown that while only 2\% of the code is spent on allocation, the overall execution time can vary up to 72\% when using different allocators~\cite{li2023nextgen} in a manner similar to butterfly effect~\cite{lorenz1963deterministic} (where a small action somewhere can cause a large effect elsewhere).
The effect is even stronger for \textbf{multi-threaded} applications.
In our experiments with three industry-grade allocators, JeMalloc (from Meta), Mimalloc (from Microsoft), and TCMalloc (from Google), 
we observe up to $2.71 \times$ overall application execution time variation in overall application execution time across the allocators, as illustrated in Figure~\ref{fig_bfs}(a) and (b) (both use 16 threads). \par

In modern \textbf{multi-threaded} multi-core systems, memory allocation is a complex task that needs to excel at many aspects, including (a) performing fast allocation and deallocation of objects, (b) handling objects of various sizes efficiently with minimal fragmentation, (c) returning freed space quickly for future allocations, (d) being scalable to many threads and cores, and so on.
Efforts to achieve fast and concurrent memory allocation while minimizing fragmentation have been ongoing since the last century.
Multiple software solutions, in the form of optimized malloc libraries, have been designed and implemented~\cite{berger2002reconsidering, evans2006scalable, pheatt2008intel, ptmalloc, tcmalloc, aigner2015fast, leijen2019mimalloc, lietar2019snmalloc, hunter2021beyond, maas2021adaptive, yang2023numalloc, zhou2024characterizing}. 
However, all modern software allocators face a common challenge: they \textit{rely on complex metadata structures}. 
These sophisticated metadata structures negatively impact overall program performance in two key ways: \textbf{cache pollution} and \textbf{metadata synchronization}.

Allocator metadata is likely to pollute processor caches, by evicting cache lines containing program user data. 
Figure~\ref{fig_bfs}(c) shows profiling results for the BFS (Breadth-first Search) workload.  We note that the conflicts with allocator metadata lead to $28.3\%$ of all cache misses in TCMalloc\footnote{We chose TCMalloc because of its wide-spread use in research and commercial datacenters, as well as its profiling utilities~\cite{kanev2017mallacc, hunter2021beyond, maas2021adaptive, zhou2024characterizing}. } when using 16 threads.
In addition to cache pollution, the cross-core metadata synchronization also limits the scalability of modern allocators. 
To support concurrent allocation in different threads, state-of-the-art allocators employ thread-local cache to maintain metadata associated with each thread~\cite{evans2006scalable, tcmalloc, leijen2019mimalloc}. 
To avoid memory \textit{blowup}\cite{berger2000hoard} (memory consumption grows linearly with the number of threads), modern allocators use a centralized cross-thread free memory pool to balance the memory consumption of each thread~\cite{leijen2019mimalloc, tcmalloc, hunter2021beyond}.
Accessing both thread-local and cross-thread metadata is expensive. The logical thread and physical core mapping are transparent to allocators, resulting in non-negligible overhead for cross-core synchronization~\cite{asgharzadeh2022free}. 
For example, in Larson~\cite{larson1998memory} (a workload models realworld server-client behaviors) from \textit{mimalloc-bench}~\cite{mimalloc-bench}, 22.8\% of overall cycles are spent on cross-core synchronization (Figure~\ref{fig_bfs}(d)). \par

Prior solutions present an opportunity to use a hardware accelerator near the core for system functions~\cite {karandikar2021hardware, karandikar2023cdpu, Wang2023Memento, kanev2017mallacc}. 
However, memory allocation in \textbf{multi-threaded} scenarios has its own challenges, especially the coherency issue (more in \S~\ref{section:background_limit}).
\textit{Instead of ``an accelerator per core'', we propose \ASNGMalloc, a system that centralizes memory allocation tasks to a support-core}. To reduce the cross-core synchronization overhead, \ASNGMalloc employs a new communication interface that routes data directly to the \SupportCore through the coherent network.
We also add hardware support, i.e., hardware message queues, in the \SupportCore to improve the allocation request throughput.
By reducing the cache conflicts and alleviating the cross-core metadata synchronization issues, \ASNGMalloc makes the CPU cycles more productive. \par

Another existing approach for reducing cache pollution and metadata synchronization (without triggering coherency issues) is to isolate the allocation functions from the rest of the application and offload the allocation tasks to a separate core in the system~\cite{tiwari2010mmt, li2023nextgen}. 
However, a full-fledged powerful core includes features that are not necessary for memory allocators (e.g., floating point arithmetic operations), resulting in energy inefficiency.
Instead of a normal core, \textit{\ASNGMalloc uses a lightweight \SupportCore for memory allocation tasks}. 
The lightweight \SupportCore is a micro-controller grade core with about $1$$\sim$$2\%$ of a modern processor chip area (the area of an Intel Xeon Max socket is $400mm^2$, while the area of a simple RISC-V core can be less than $4mm^2$~\cite{wang202430}). \par

Results show that on \textbf{multi-threaded} workloads, \ASNGMalloc achieves an average of \JeSpeedup, \TCSpeedup, and \MiSpeedup\space overall performance speedup over Jemalloc, TCMalloc, and Mimalloc, respectively.
As a result of using a lightweight \SupportCore, \ASNGMalloc also achieves \JeEnergy, \TCEnergy, and \MiEnergy\space energy savings over the three industry allocators. 
We also compare \ASNGMalloc with two hardware allocator accelerators \textit{Mallacc}~\cite{kanev2017mallacc} and \textit{Memento}~\cite{Wang2023Memento}. 
\ASNGMalloc achieves \MaSpeedup\space and \MeSpeedup\space  overall performance speedups over \textit{Mallacc} and \textit{Memento} on multi-threaded workloads, respectively. \par

%% file: 02_background.tex
This section includes an overview of memory allocators (\S~\ref {section:background_uma}) and their performance challenges with multi-threaded workloads (\S~\ref{section:background_uma_multi_thread}). 
We also describe the limitations of current hardware accelerators for multi-threaded applications (\S~\ref{section:background_limit}) and motivate the need for \ASNGMalloc (\S~\ref{section:background_scmall}). \par

\subsection{Background of Memory Allocators}
\label{section:background_uma}
Memory allocation exists at two levels: user level and kernel level. 
A user-level memory allocator (UMA, referred to as memory allocator in this paper) implements user-level memory allocation tasks, such as ptmalloc2~\cite{ptmalloc} (default Glibc allocator), TCMalloc~\cite{hunter2021beyond}, and Mimalloc~\cite{leijen2019mimalloc}. 
For kernel-level operations, UMAs use $mmap()$\footnote{In this paper, we will focus on the user-level memory management and use the default $mmap()$ system call. }
system calls to make operating systems map memory segments for each process. 
\par

We show the pseudo-code of $malloc()$ (UMA memory allocation interface) in Code~\ref{alg_malloc}.
To minimize the overhead of $mmap()$ system calls, most UMAs request large chunks of memory at a time and allocate memory out of these chunks to satisfy allocation requests (known as the \textit{malloc\_fast}).
UMAs only make additional $mmap()$ calls when pre-allocated space is exhausted (known as the \textit{malloc\_generic}).
With the help of \textit{malloc\_fast}, state-of-the-art memory allocators spend less than 10\% of the allocation cycles in the kernel space~\cite{kanev2017mallacc, zhou2024characterizing}.
Managing the pre-allocated memory in a way that strikes \textit{a balance between allocation speed and memory fragmentation} remains a challenging problem, drawing research attention from both software and hardware communities.
\par

\lstset{
  basicstyle=\ttfamily\small,
  keywordstyle=\color{blue},
  commentstyle=\color{gray},
  stringstyle=\color{red},
  numbers=left,
  numberstyle=\tiny,
  breaklines=true,
  frame=single,
  captionpos=b
}

\begin{figure}[ht]
\centering
\begin{minipage}{0.7\linewidth}
\centering
\lstset{language=C}
\begin{lstlisting}[caption={Malloc Pseudo Code}, label={alg_malloc}]
void* malloc(size_t size) {
    void *memory_block = malloc_fast(size);
    if (memory_block == NULL) return malloc_generic(size);
    else return memory_block;
}
void* malloc_fast(size_t size) {
    get the first memory_block from the free block list;
    return memory_block;
}
void* malloc_generic(size_t size) {
    acquire memory pages from the OS (mmap syscall);
    update/initialize free block list;
    return malloc_fast(size);
}
\end{lstlisting}
\end{minipage}
\end{figure}

\subsection{Challenges in Multi-threaded UMAs}
\label{section:background_uma_multi_thread}
Modern UMAs support concurrent memory allocation and deallocation for multi-threaded applications. 
For example, TCMalloc~\cite{tcmalloc} assigns each thread/core a thread-local cache. 
The majority of allocations are satisfied from the thread-local cache on each core, instead of processed by a central allocator in series. \par

\begin{figure}[t]
    \centerline{\includegraphics[width=0.6\columnwidth,trim = 82mm 65mm 97mm 59mm, page=6, clip=true]{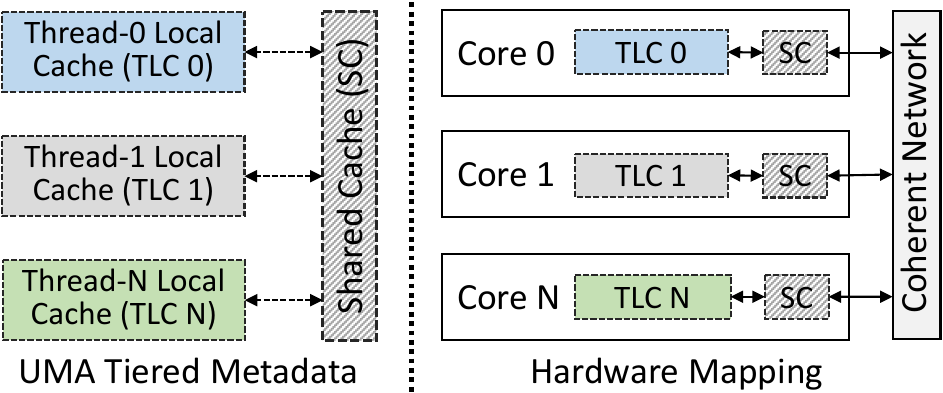}}
	\caption{An abstraction of how allocator metadata is mapped to each physical core in modern allocators. Shared Cache synchronization requires cross-core communication. }
	\label{fig_tcmalloc_metadata}
\end{figure}

One major challenge for memory allocators in handling multi-threaded applications is memory \textit{blowup}, i.e., memory consumption can grow linearly with the number of threads/cores (server-client model)~\cite{berger2000hoard}.
To address the \textit{blowup} issue, state-of-the-art allocators~\cite{leijen2019mimalloc, tcmalloc} employ tiered metadata.
As shown in Figure~\ref{fig_tcmalloc_metadata}, allocators use thread-local cache (thread-local metadata) to support fast and small concurrent memory allocation in each thread.
If the thread-local cache is exhausted, the allocation request will search for free memory space of other threads through the shared cache (inter-thread shared metadata). \par

Though such tiered metadata can help with the \textit{blowup} issue, it makes the system suffer from (a) additional cache pollution, and (b) synchronization overhead. \par
\noindent{\textbf{Cache pollution:}}
Accessing the metadata can pollute the CPU caches, by evicting cache lines containing user data~\cite{li2023nextgen}. 
In multi-threaded allocators, one allocation request can result in: (a) accessing the thread-local cache; (b) accessing the shared cache if the thread-local cache is exhausted, and potentially the thread-local cache of other threads for additional free space.
Cache pollution poses an even greater challenge in multi-threaded applications since allocator metadata can even pollute cache lines of other cores (right part of Figure~\ref{fig_tcmalloc_metadata}). \par

\noindent{\textbf{Synchronization:}}
Modern allocators use intensive synchronization logic, e.g., atomic instructions~\cite{asgharzadeh2022free}, to secure the metadata coherency (for both thread-local and shared cache). 
Cross-core synchronization can downgrade the system performance since the communication overhead increases with the number of cores in the system~\cite{schweizer2015evaluating, asgharzadeh2022free}. \par

\subsection{Current Hardware Solutions are Limited}
\label{section:background_limit}
Offloading system functions (including memory allocators) to accelerators near the core have been proposed in the past~\cite{rezaei2006intelligent, kanev2017mallacc, karandikar2021hardware, karandikar2023cdpu, Wang2023Memento}. 
However, ``an accelerator per core'' solution falls short for multi-threaded systems. 
As shown in Figure~\ref{fig_tcmalloc_metadata} (right side), the tiered metadata of one thread can exist in multiple cores.
Implementing a coherent protocol between all processors and all accelerators is expensive.
An alternative solution is to make allocators not use shared cache and use thread-local cache only~\cite{Wang2023Memento}; however, this leaves the \textit{blowup} issue unresolved. \par

Another limitation of accelerators is that accelerators are not flexible to software challenges. For example, TCMalloc has an average of $1.33$ new commits every day~\cite{tcmalloc}. Regarding LOC (line of codes), last year, the most active author made $32.2$ LOC changes on average per day. \par

\begin{table}[t]
\caption{SOTA software and hardware memory allocators. }
\label{table_related}
\centering
\setlength{\tabcolsep}{1pt}
\renewcommand{\arraystretch}{1}
\begin{tabular}{| >{\centering\arraybackslash}p{3cm} |
                   >{\centering\arraybackslash}p{5cm} |
                   >{\centering\arraybackslash}p{2cm} 
                   |}
 \hline
\textbf{Work} & \textbf{Weakness/Feature} & \textbf{Multi-Thread} \\ 
\hline \hline 
Jemalloc~\cite{evans2006scalable} & \multirow{3}{*}{\shortstack[l] {Cache pollution; Cross-core \\ metadata synchronization overhead }} 
& Yes \\
\cline{1-1} \cline{3-3}
TCMalloc~\cite{tcmalloc} & & Yes  \\
\cline{1-1} \cline{3-3}
Mimalloc~\cite{leijen2019mimalloc} & & Yes  \\
\hline
Mallacc~\cite{kanev2017mallacc} & 
\multirow{2}{*}{\shortstack[l] {Additional \textbf{dedicated} hardware; \\ \textit{blowup} issue in multi-threaded apps }}
& No  \\
\cline{1-1} \cline{3-3}
Memento~\cite{Wang2023Memento} & & Limited  \\
\hline
\ASNGMalloc (Ours) & Additional lightweight co-processor & \textbf{Yes}  \\
\hline

\end{tabular}
\end{table}

\subsection{A Support-Core with Efficient Data Transfer}
\label{section:background_scmall}

Instead of ``an accelerator per core'', a centralized solution, using a \SupportCore for all memory allocation tasks can help alleviate cache pollution and metadata coherency issues. 
Table~\ref{table_related} summarizes state-of-the-art software and hardware memory allocators. 
General-purpose\footnote{There are also other profiling and machine learning guided software allocators~\cite{maas2020asplos, oh2021lctes}, which require offline profiling/training with additional runtime overhead.
These methodologies are orthogonal to our approach in \ASNGMalloc. } software allocators~\cite{evans2006scalable, tcmalloc, leijen2019mimalloc} suffer from cache pollution and synchronization overhead. 
Current hardware allocators~\cite{kanev2017mallacc, Wang2023Memento} have limited support for multi-threaded applications. \par

While prior prototypes of offloading memory allocator to a separate core achieve performance gains for single-threaded applications~\cite{tiwari2010mmt, li2023nextgen}, their performance remains constrained by the substantial overhead of cross-core communication (more on this in \S~\ref{section_idle_core}). 
For example, a single atomic instruction (required for inter-core synchronization) can consume up to 700 cycles in modern processors~\cite{asgharzadeh2022free}, while most allocation functions can be finished within 100 cycles~\cite{kanev2017mallacc, zhou2024characterizing}.
A more desired and feasible direction is to offload memory allocators to a \SupportCore with software and hardware support for efficient cross-core communication. 
In addition, compared with harvesting one of the primary processing cores, a \textbf{lightweight} \SupportCore is adequate for memory allocation tasks, as they do not need floating-point and vector operations or complex control logic.
Following this insight, we develop \ASNGMalloc. \par

%% file: 03_overview.tex
\begin{figure}[t]
    \centerline{\includegraphics[width=0.5\columnwidth,trim = 71mm 177mm 73mm 66mm, clip=true, page=1]{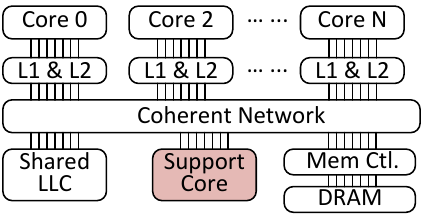}}
	\caption{\ASNGMalloc overall architecture (\SupportCore connects to the coherent network). }
	\label{fig_overall}
\end{figure}

This section provides an overview of \ASNGMalloc. As shown in Figure~\ref{fig_overall}, we add a \textbf{lightweight} \SupportCore connected to the coherent network.
The \textit{support-core} only needs an L1 cache to hold allocator metadata.
The \textit{support-core} does not require any floating point or vector units for allocation tasks.
In the following sections, we elaborate on the design of \ASNGMalloc. 
We develop a lightweight interface between main cores and the \SupportCore for offloading memory allocators (\S~\ref{section_control}). This reduces communication costs compared to offloading memory allocators to an idle core.
We discuss the runtime of executing allocation tasks in the \SupportCore (\S~\ref{section_data_transfer}). 
\ASNGMalloc uses both software and hardware approaches to reduce cache pollution and contention during allocation runtime.
\par

%% file: 04_control.tex
In a conventional system, offloading a function to a separate core typically requires expensive cross-core synchronization schemes, e.g., atomic instructions like exclusive read/write.
\ASNGMalloc employs a lightweight function invocation interface by using signals and data packets (\S~\ref{section_signal}) with ISA and compiler support (\S~\ref{section_isa_support} and \S~\ref{section_compiler_support}). 
Data packets can route to the \SupportCore directly, without triggering additional cache snoop requests. \par

\subsection{Signal and Data Packets}
\label{section_signal}
\begin{figure*}[t]
	\centerline{\includegraphics[width=\columnwidth,trim = 1mm 95mm 60mm 21mm, clip=true, page=2]{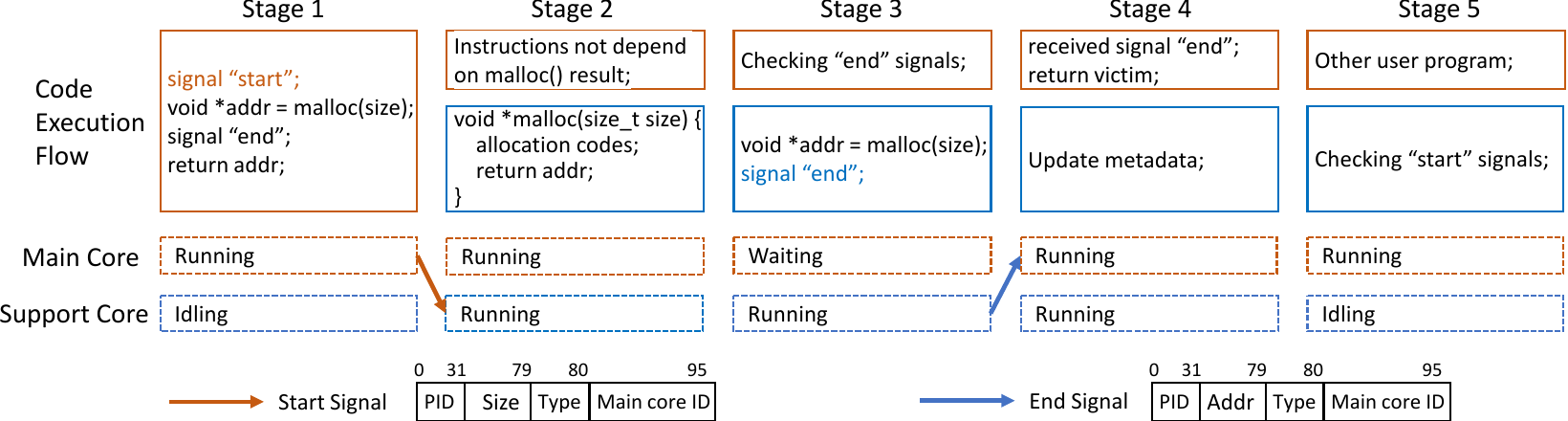}}
	\caption{Top: Invocation interface in \ASNGMalloc. Blue indicates codes executed on \SupportCore. Bottom: Signal data packet format. }
	\label{fig_signal}
\end{figure*}

Figure~\ref{fig_signal} shows the allocation function (\SupportCore) invocation interface in \ASNGMalloc (using $malloc()$ as an example). 
The allocation function parameters (allocation request size) and process identification are transferred along with the signals. 
The right side of Figure~\ref{fig_signal} shows the data packet format for this purpose (more details about the data packet itself will be discussed in \S~\ref{section_hardware}). \par

The $malloc()$ process is broken down into 5 stages.
Once the main core reaches the beginning of the $malloc()$ function, it will send the ``start'' signal to the \SupportCore (Stage 1).
Along with the ``start'' signal, the data packet is also transferred to the \SupportCore. 
After receiving the ``start'' signal from the main cores, the \SupportCore will start executing the $malloc()$ function (Stage 2). 
The main core that calls the \SupportCore will skip the $malloc()$ execution. 
Meanwhile, the main core can execute instructions that do not depend on $malloc()$ return value. 
If no such instructions are available, the main core will enter the next stage (Stage 3) and wait for the ``end'' signal from the \SupportCore. 
\ASNGMalloc masks the exception and interrupt handler at this stage to guarantee the main core can receive the ``end'' signal.
When the \SupportCore reaches the end of the $malloc()$ function, it will send an ``end'' signal along with the returned allocated address to the main core to wake the core waiting for $malloc()$.
At Stage 4, main cores can execute instructions that rely on the $malloc()$ return value. 
The \SupportCore will update the allocator metadata asynchronously, removing this operation from the critical application execution path.  
Once the metadata is updated, the \SupportCore will switch to Stage 5, awaiting the next ``start'' signal. \par

One prerequisite for the invocation interface is that signals must be aware of the program context information, e.g., program counter, of the allocation functions. 
Such information is architecture and compiler dependent, necessitating additional ISA (\S~\ref{section_isa_support}) and compiler (\S~\ref{section_compiler_support}) support. \par

\subsection{ISA Support}
\label{section_isa_support}
Main cores will send ``start'' signals to the \SupportCore for each allocation request.
Similarly, the \SupportCore will send ``end'' signals back once finishing allocation.
To support such signals, \ASNGMalloc enhances current ISAs by adding 4 new unconditional branch instructions\footnote{Reusing existing instructions by utilizing reserved bits is also possible. We demonstrate using new instructions in this paper.}. \par

We define two new instructions for main cores, ``mallocstart()'' and ``freestart()''. 
When executing these instructions, main cores will send the ``start'' signal to the \SupportCore for both instructions. 
When committing these instructions, ``mallocstart()'' will check for the ``end'' signal and update the destination register value (return value of the $malloc()$ function).
Since memory deallocation is not on the critical path, \ASNGMalloc supports processing $free()$ functions asynchronously instead of blocking the execution of the main cores. 
Therefore, the ``freestart()" instruction can retire without waiting for an end signal from the \SupportCore. \par

We also define two new instructions for the \SupportCore, ``mallocend()'' and ``freeend()''.
These two instructions are inserted at the end of malloc() and free() functions.
These two instructions indicate the end of one allocation/deallocation function and will set  the PC (program counter) to the start address of the following pending $malloc()$/$free()$ request if there are any.
Otherwise, the \SupportCore will stall the pipeline to save energy. 
The ``mallocend()'' will send the ``end'' signal and the destination register value to the main core, while ``freeend()'' retires as normal instructions. \par

\begin{figure}[t]
	\centerline{\includegraphics[width=0.7\columnwidth,trim = 122mm 90mm 49mm 80mm, clip=true, page=7]{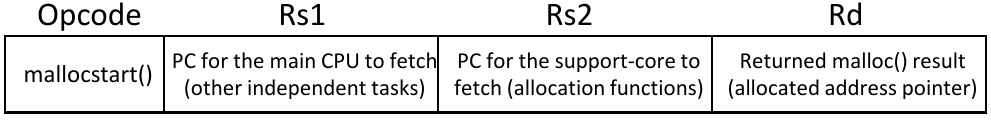}}
	\caption{Instruction format (using \textit{mallocstart} as an example).}
	\label{fig_ins}
\end{figure}

\subsection{Compiler Support}
\label{section_compiler_support}
With the new instructions in \ASNGMalloc, the main cores will have the information of when to branch to the \SupportCore execution.
However, where to branch is compiler-dependent.
This part will discuss the compiler support for the added instructions in \ASNGMalloc. \par

When the main cores issue ``mallocstart()'' (or ``freestart()'') instructions, the \SupportCore will start fetching instructions related to memory allocation tasks.
Where to fetch the instructions in the \SupportCore, i.e., the destination PC of the branch instructions, should be part of the instruction and generated during compilation time (instruction example shown in Figure~\ref{fig_ins}).
To solve this, we add a post-compilation pass\footnote{Instead of a post-compilation pass, this process can also be integrated into the current compiler, e.g., by modifying gcc.} to get such information. 
We pack \ASNGMalloc as either a static or shared library. 
Programmers can choose whether to use \ASNGMalloc or other allocators during the program linking stage, without changing program source codes.

%% file: 05_data.tex
In addition to the allocator invocation, \ASNGMalloc provides additional support for allocation runtime. 
\ASNGMalloc reduces the cache pollution and contention by using both software approaches (\S~\ref{section_software_approach}). 
\ASNGMalloc handles contention well by additional hardware support in the \SupportCore (\S~\ref{section_hardware}). \par

\subsection{Software Approach}
\label{section_software_approach}
\ASNGMalloc employs a segregated metadata layout, making data accessed by the \SupportCore separate from other data accessed by main cores, thus reducing metadata cache pollution 
 (\S~\ref{section_software_coherency}) and contention (\S~\ref{section_contention}).

\subsubsection{Reduce Cache Pollution}
\label{section_software_coherency}
During allocation runtime, accessing the same memory location from both main cores and \SupportCore can result in significant performance overhead due to cache pollution. 
To avoid such behavior, \ASNGMalloc employs a segregated metadata layout, making data accessed by the \SupportCore separate from other data accessed by main cores. \par

\begin{figure}[t]
	\centerline{\includegraphics[width=0.54\columnwidth,trim = 96mm 81mm 92mm 53mm, clip=true, page=5]{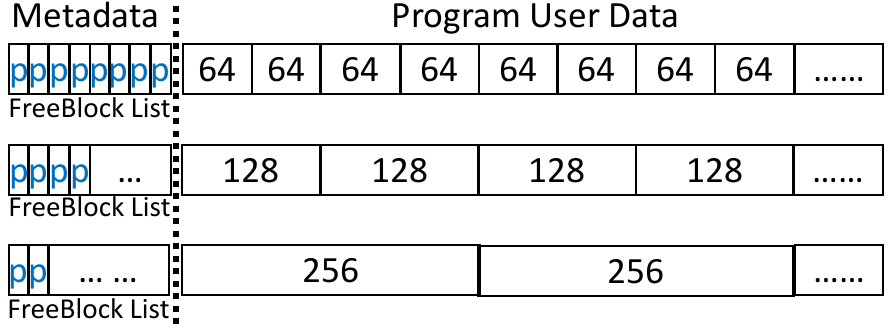}}
	\caption{Segregated metadata layout in \ASNGMalloc. Metadata (left side) contains linked lists of pointers for available free blocks. }
	\label{fig_layout}
\end{figure}

As shown in Figure~\ref{fig_layout}, \ASNGMalloc maintains linked lists as metadata to indicate available free space with different sizes.
During memory allocation, the \SupportCore will only access the metadata and get the pointer of the next available memory block, i.e., the 64/128/256-byte\footnote{64/128/256 are block size examples. Design methodologies are well explored in software allocators~\cite{leijen2019mimalloc, tcmalloc} and are not the focus of \ASNGMalloc.} 
block in the right part of Figure~\ref{fig_layout}.
During the allocation phase, the \SupportCore can make $mmap()$ system calls (if necessary) to claim memory from the OS, making the majority of system memory metadata accessed exclusively by the \SupportCore.
After the \SupportCore finishes allocation tasks, the main cores will receive the address pointer for the allocated memory.
Main cores will only access the allocated memory blocks, which are segregated from allocator metadata, which resolves the cache pollution issue in other software allocators. \par

\subsubsection{Address Metadata Contention}
\label{section_contention}
In the previous discussion of the invocation interface in \ASNGMalloc, assuming only one main core interacts with the \SupportCore. 
However, metadata contention occurs when multiple cores simultaneously attempt to access allocator metadata (e.g., shared cache), which is a common issue in memory allocation for multi-threaded applications.
Software solutions use atomic instructions to serialize changes made to the allocator metadata (Figure~\ref{fig_tcmalloc_metadata}) to avoid race conditions; however, the overhead of these atomic instructions is high. \par

With the segregated metadata layout in \ASNGMalloc, all allocation requests are processed by the \SupportCore, eliminating such metadata contention. 
All allocator metadata is physically stored within the \SupportCore, preventing inter-core metadata exchanges. 
This design eliminates the need for atomic instructions, unlike other software allocators.
In addition, such a design in \ASNGMalloc can also eliminate the \textit{blowup} issue and reduce the design complexity of the allocator itself. \par

\subsection{Hardware Approach}
\label{section_hardware}
\ASNGMalloc eliminates metadata contention that exists in modern allocators but comes at the cost of hardware contention.
Hardware contention occurs when multiple cores send allocation requests to the \SupportCore simultaneously: the \SupportCore has to process these requests serially. \par

\begin{figure}[t]
	\centerline{\includegraphics[width=0.6\columnwidth, trim = 82mm 62mm 140mm 54mm, clip=true, page=4]{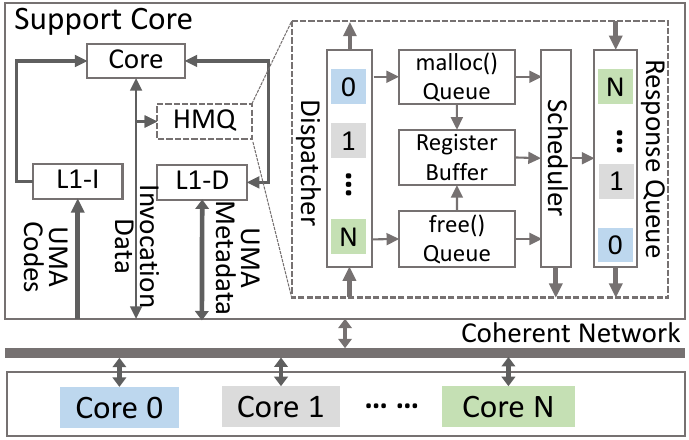}}
	\caption{Hardware message queues (HMQ) added in the \SupportCore to address contention and improve allocation throughput.
 }
	\label{fig_message_queue}
\end{figure}

\ASNGMalloc relieves the hardware contention by using additional hardware message queues in the \SupportCore (HMQ shown in Figure~\ref{fig_message_queue}). 
When the \SupportCore receives a memory allocation request from main cores, the dispatcher first pushes the request to the $malloc()$ queue or $free()$ queue\footnote{When the $malloc()$/$free()$ queue is full, additional requests could be buffered using reserved physical memory~\cite{margaritov2021ptemagnet} and fetched to the queue once space becomes available.}, depending on the type of the request. The scheduler will pop the first element from the $malloc()$ queue if it is not empty, or the first element from the $free()$ queue if the $malloc()$ queue is empty. The scheduler prioritizes $malloc()$ requests since it is on the (performance) critical path.
In addition, the scheduler serves allocation requests from different cores in a round-robin fashion (main core ID is part of the data packet shown in Figure~\ref{fig_signal}), which gives each main core fair access to the \SupportCore.
After finishing the execution of $malloc()$/$free()$ on the \SupportCore, the returned results will be pushed to the response queue (if it has). 
The \SupportCore will then pop the first element from the response queue and send the ``end'' signal and the return value (if it has) back to the main core. \par

In addition to the two message queues, we add one register buffer in the \SupportCore.
The register buffer is implemented as a directly mapped cache.
It uses the process ID as the index and stores required system control and segment registers for address translation.
Instead of transferring these system registers in each data packet (Figure~\ref{fig_signal}) on the fly, the main core will only send them to the \SupportCore in its first allocation request (transferring data packets can be implemented as writing to the device memory).
This also helps achieve a faster allocation invocation in \ASNGMalloc. \par

%% file: 08_evaluation.tex
We demonstrate \ASNGMalloc’s effectiveness by answering the following questions through experimental evaluation:
\begin{itemize}
    \setlength\itemsep{0em}
    \item Can \ASNGMalloc improve overall program performance when compared with other allocators? If so, what are the sources of the performance gains? (\S~\ref{section_performance})
    \item Can \ASNGMalloc achieve energy savings given the added  \SupportCore in the system? (\S~\ref{section_energy})
    \item Is \ASNGMalloc necessary? Can the same performance gains be achieved using existing hardware, e.g., partitioning cache or harvesting an idle core? (\S~\ref{section_software})
    \item What is the scalability of \ASNGMalloc? (\S~\ref{section_discussion})
\end{itemize}

Highlights of our results are: 
\begin{itemize}
    \setlength\itemsep{0em}
    \item \ASNGMalloc achieves \JeSpeedup, \TCSpeedup, \MiSpeedup, \MaSpeedup, and \MeSpeedup \space speedups over Jemalloc, TCMalloc, Mimalloc, \textit{Mallacc}, and \textit{Memento+} when using 16 threads. The gains are from the reduction of cache pollution and the elimination of metadata synchronization.
    \item \ASNGMalloc achieves \JeEnergy, \TCEnergy, \MiEnergy, \MaEnergy, and \MeEnergy \space  energy savings over Jemalloc, TCMalloc, Mimalloc, \textit{Mallacc}, and \textit{Memento+} on the 16-core system (running 16 threads). 
    The reason is that \ASNGMalloc uses a lightweight \SupportCore, which consumes only $\sim$$2\%$ of the system power.
\end{itemize}

\subsection{Experimental Setup}
\input{08_01_evaluation}

\subsection{Does \ASNGMalloc achieve performance gains?}
\label{section_performance}
\input{08_02_evaluation}

\subsection{Does \ASNGMalloc achieve energy savings?}
\label{section_energy}
\input{08_03_evaluation}

\subsection{\ASNGMalloc vs. Using Existing Hardware}
\label{section_software}
\input{08_04_evaluation}

\subsection{Scalability of \ASNGMalloc}
\label{section_discussion}
\input{08_06_evaluation}

%% file: 08_01_evaluation.tex
\subsubsection{End-to-End Simulation Platforms}
We evaluate \ASNGMalloc using \textit{gem5}~\cite{binkert2011gem5, lowe2020gem5} in full system mode\footnote{We are not able to prototype \textit{SpeedMalloc} in a real system, e.g., offloading to an FPGA, as the \textit{support-core} has to be connected to the coherent bus above the last-level-cache level to achieve a low communication overhead, which requires additional ISA and hardware not exist in current systems.}.
The detailed configuration of the hardware simulated is shown in Table~\ref{tab:gem5_settings}. 
We use the OoO (Out-of-Order) processor model for the big core and the minor (in-order) processor model for the little core~\cite{gem5cpu}.
We use GCC 9.4 as the compiler with -O3 optimization to compile \ASNGMalloc and all workloads. 
We also use the ``-march=armv8.1-a'' optimization to support far atomic instruction in recent aarch64 processors.
We use McPAT~\cite{li2009mcpat} to obtain area, power, and energy values. \par\

\subsubsection{Benchmarks}
We evaluate \ASNGMalloc on both single-threaded and multi-threaded workloads. The workloads are from 
\textit{Mimalloc-bench}~\cite{mimalloc-bench} (Mimalloc's evaluate suite~\cite{leijen2019mimalloc}\footnote{Jemalloc's and TCMalloc's benchmark suites are not publicly available.}), \textit{GAPBS}~\cite{beamer2015gap} (graph processing), \textit{Redis-benchmark}~\cite{redis-benchmark} (in-memory database), and \textit{NAS}~\cite{bailey1991parallel} (computational fluid dynamics applications). 
We simulated the \textbf{end-to-end}\footnote{Simulating a subset using simpoints~\cite{sherwood2002automatically} would not work since \ASNGMalloc needs to analyze how allocator impacts on the whole program performance. 
This makes it impractical to simulate large workloads, such as SPEC benchmarks, which are also unsuitable for memory allocation studies~\cite{zhou2024characterizing}.
In addition, most of the SPEC benchmarks do not actively allocate objects in a stable state, which makes them unsuitable for studying memory allocation~\cite{zhou2024characterizing}.
} performance of each workload, with instruction counts from 47 million to 7 billion. 
\par

\begin{table}[t]
    \caption{\textit{gem5} simulator hardware configuration. }
    \label{tab:gem5_settings}
    \centering
    \setlength{\tabcolsep}{3pt}
    \renewcommand{\arraystretch}{1}
    
    \begin{tabular}{| >{\centering\arraybackslash}p{2cm} |
                    >{\centering\arraybackslash}p{2cm} |
                   >{\centering\arraybackslash}p{7cm} |}
        \hline
       \multirow{5}{*}{\shortstack[l] {\textbf{Main cores}}} & \textbf{Core} & 1-16 $\times$ \textit{AArch64} big core \\
        \cline{2-3}
        & \multirow{3}{*}{{\textbf{L1\&L2 Cache}}} & 32KB 8-way L1d; 4 cycles latency \\
        \cline{3-3}
        & & 32KB 8-way L1i; 4 cycles latency \\
        \cline{3-3}
        & & 256KB 8-way L2; 12 cycles latency \\
        \cline{2-3}
        & \textbf{LLC} & 2MB per Core 16-way; 24 cycles latency \\
        \cline{1-3}
        \multirow{4}{*}{\shortstack[l] {\textbf{Support-Core}}} & \multirow{1}{*}{\shortstack[l] {\textbf{Core}}} & 1 $\times$ \textit{AArch64} little core \\
        \cline{2-3}
        & \textbf{HMQ} & 128-entry dispatch queue \& response queue \\
        \cline{2-3}
        & \multirow{2}{*}{\textbf{L1 Cache}} & 16KB 4-way L1d; 2 cycles latency \\
        \cline{3-3}
        & & 16KB 4-way L1i; 2 cycles latency \\
        \cline{1-3}
        \multicolumn{2}{|c|}{\textbf{Main-Support Core Latency}}
        & 
        8 cycles latency\tablefootnote{The low-latency can be achieved by bypassing the main core cache without triggering the cache line indexing, which has been applied in prior works~\cite{wang2016caf, wu2021virtual} (more studies in \S~\ref{section_idle_core}).} \\
        \cline{1-3}
        \multicolumn{2}{|c|}{\multirow{2}{*}{\textbf{DRAM}}} & 16GB 2400MHz DDR4, 8 devices/rank, 2 ranks/channel, tCAS=tRCD=tRP=14.16ns, tRAS =32ns \\
        \hline
    \end{tabular}
\end{table}

\noindent\textbf{Single-threaded workloads.}
We use two single-threaded workloads \textsl{Espresso}~\cite{grunwald1993improving} and \textsl{Cfrac}~\cite{cfrac} from \textit{Mimalloc-bench}.
\textsl{Espresso}~\cite{grunwald1993improving} is a programmable logic array analyzer, in the context of cache-aware memory allocation.
\textsl{Cfrac}~\cite{cfrac} is an implementation of continued fraction factorization, using many small short-lived allocations.
We use 6 workloads from Redis-benchmark~\cite{redis-benchmark}: \textsl{LPUSH}, \textsl{RPUSH}, \textsl{LPOP}, \textsl{RPOP}, \textsl{SADD}, and \textsl{SPOP}. \par

\noindent\textbf{Multi-threaded workloads.}
We selected 7 representative multi-threaded workloads from \textit{Mimalloc-bench}, each exhibiting a different allocation characteristic and showing significant performance differences when using different allocators~\cite{leijen2019mimalloc}.
\textsl{Larson}~\cite{larson1998memory, dang2022nvalloc} simulates realworld server-client behaviors. 
\textsl{Xmalloc}~\cite{boreham2000malloc} simulates the scenario where objects allocated by one thread are freed by another thread.
\textsl{Cache-Scratch}~\cite{berger2000hoard} (abbreviated as \textsl{Scratch}) is used to test passive-false sharing of cache lines. 
\textsl{Sh6(8)bench}~\cite{shbench} is a stress test for allocating small objects more frequently. 
\textsl{Mstress}~\cite{mstress} simulates real-world server-like allocation patterns, where objects can migrate between threads with lifetimes. 
\textsl{Alloctest}~\cite{alloc-test} simulates intensive allocation workloads with a Pareto size distribution. \par

We selected 2 widely-used graph algorithms from \textit{GAPBS}: Breadth First Search (\textsl{BFS}) and Betweenness Centrality (\textsl{BC}), using a synthetic graph with 64k nodes with an average degree of 100.
These two workloads use queue data structures that would allocate/deallocate the heap memory frequently, while other workloads, e.g., DFS (Depth First Search) based algorithms, use stack memory through recursive function calls (stack memory grows automatically instead of being managed by memory allocators). \par

We selected Data Cube~\cite{frumkin2003arithmetic} (\textsl{DC}) from \textit{NAS}, which benchmarks grid capabilities to handle the data movement across large distributed data sets. \textsl{DC} is also the only workload that allocates heap memory during runtime in the \textit{NAS} suite. \par

\noindent\textbf{Workloads are representative.}
Table.~\ref{table_butter_fly} shows the percentage of allocation instruction of the overall program for the benchmarks. 
For the 10 multi-threaded workloads we used, on average \Allocation \space of instructions is spent on memory allocation tasks.
In addition, there is a large variance (from $0.37\%$ to $7.22\%$) of the overall instructions on memory allocation. Such variance indicates these workloads are highly representative and are able to cover both allocation intensive and non-intensive workloads. 
The suite we select is a good proxy for realworld workloads, which spans the full gamut of memory allocation overheads (from $2.7\%$ to $10.7\%$ and an average of $\sim$$5\%$ of overall execution time is spent on memory allocation in Google datacenter workloads~\cite{hunter2021beyond}). 
\par

\subsubsection{Baselines}
We compare \ASNGMalloc with three state-of-the-art software solutions and two hardware solutions. \par

\noindent\textbf{Software baselines.}
Our software baselines are Jemalloc from Meta~\cite{evans2006scalable}, TCMalloc from Google~\cite{tcmalloc}, and Mimalloc from Microsoft~\cite{leijen2019mimalloc}. 
For a fair comparison, we use the default hyperparameter (of each allocator) for all workloads.
These state-of-the-art UMAs are chosen as the baselines for comparison as they are highly optimized for industry-scale warehouse systems. \par

\noindent\textbf{Hardware baselines.}
There are other hardware accelerators for \textbf{single-threaded} memory allocation, e.g., \textit{Mallacc}~\cite{kanev2017mallacc} and \textit{Memento}~\cite{Wang2023Memento}.
\textit{Mallacc} was built atop of TCMalloc\footnote{It is infeasible to build \textit{Mallacc} atop Mimalloc, which uses aggregated metadata layout (address space of metadata and user data are intertwined)~\cite{li2023nextgen}.} and the key idea of \textit{Mallacc} is to have a dedicated \textit{malloc} cache at the L1-cache level. 
We implemented \textit{Mallacc} by adding a 16KB \textit{malloc} cache.
\textit{Memento} achieved further performance improvement by implementing an accelerator near the core (named \textit{Object Allocator}) dedicated to allocation tasks.
To support multi-threaded systems and avoid memory \textit{blowup} (details in \S~\ref{section:background_limit}) in the original \textit{Memento} design, we implemented \textit{Memento+} by adopting the shared cache (transfer cache) design of TCMalloc for cross-thread communication and connecting both the \textit{Object Allocator} and \textit{malloc} cache to the coherent bus.
We set the \textit{Object Allocator} latency in \textit{Memento+} the same as \textit{Memento} (each allocation request takes the same time as a single cache access if it is a \textit{malloc} cache hit~\cite{Wang2023Memento}).

%% file: 08_02_evaluation.tex
\begin{figure}[t]
\centerline{\includegraphics[width=0.9\columnwidth,trim = 2mm 3mm 7mm 4mm, clip=true]{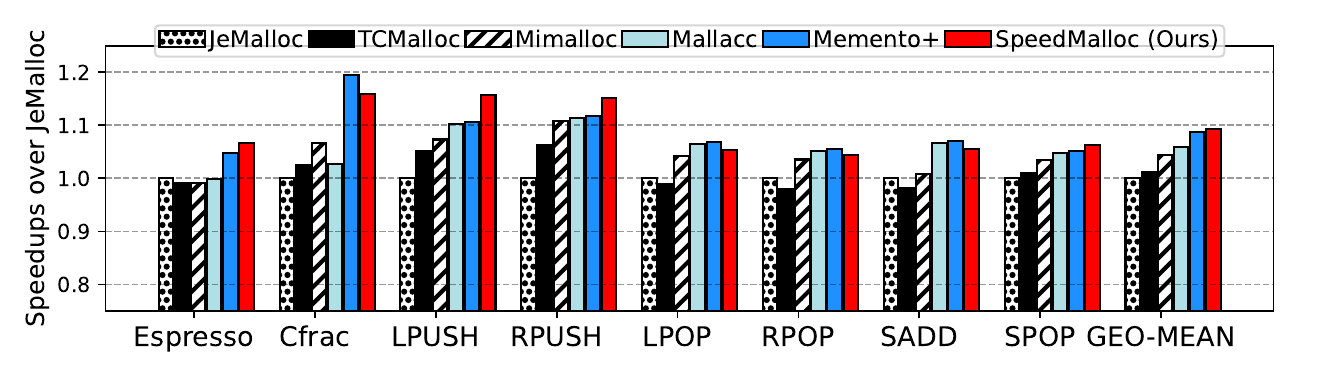}}
	\caption{Performance comparison on single-threaded workloads (the larger the better). \ASNGMalloc excels over other allocators. }
	\label{fig_performance_single}
\end{figure}

\begin{figure*}[t]
 \centerline{\includegraphics[width=\columnwidth,trim = 55mm 27mm 55mm 17mm, clip=true]{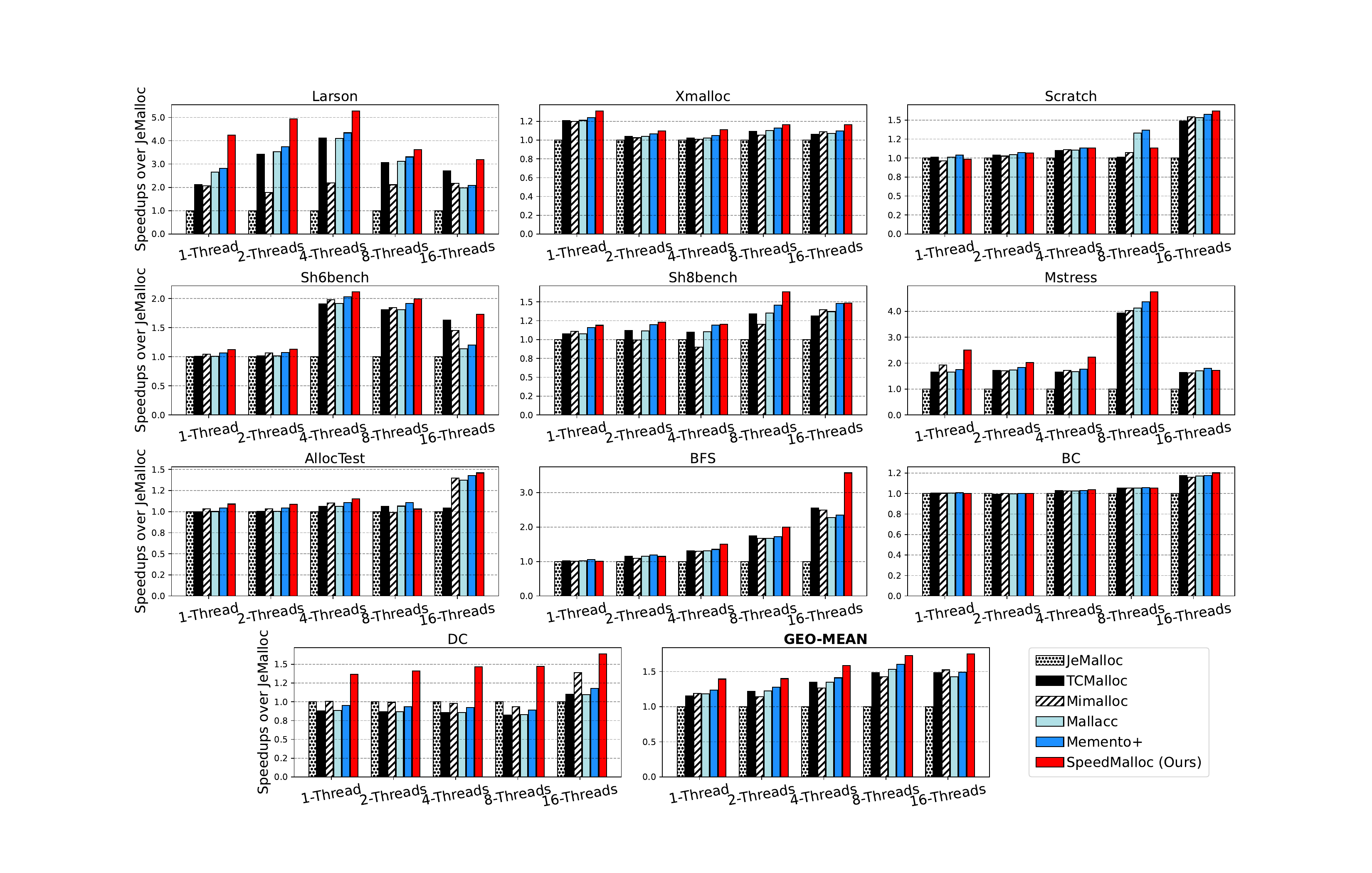}}
	\caption{Performance comparison on multi-threaded workloads (the larger the better). \ASNGMalloc excels over other allocators. }
	\label{fig_performance}
\end{figure*}

\subsubsection{Results} 
\label{section_result}

\noindent\textbf{\ASNGMalloc performs well on single-threaded workloads.}
As shown in Figure~\ref{fig_performance_single}, \ASNGMalloc achieves an average of $1.09 \times$, $1.08 \times$, $1.04 \times$, and $1.03 \times$ speedups over Jemalloc, TCMalloc, Mimalloc, and \textit{Mallacc}.
\ASNGMalloc outperforms the three software allocators by reducing cache pollution. 
When compared with Jemalloc, TCMalloc, and Mimalloc, \ASNGMalloc achieves $19.93\%$, $14.69\%$, and $13.65\%$ reduction in L2 miss cycles. \par

\textit{Memento+} uses a dedicated accelerator for fast allocation (allocation cycles contribute $\sim$$3\%$ of overall cycles for the 8 single-threaded workloads) but only buffers 16 entries (based on serverless object life cycles) of each allocation size.
The allocation size of \textit{Redis-benchmark} is more unified, making \ASNGMalloc perform better due to adopting general software allocators (e.g., Mimalloc can buffer at most 8k entries for the 8-byte allocation size). 
In general, \ASNGMalloc achieves similar performance as \textit{Memento+}\footnote{\textit{Memento+} outperforms Jemalloc by  $\sim$$1.1 \times$ in \textit{Redis-benchmark}. This is consistent with \textit{Memento} results~\cite{Wang2023Memento}.}. \par

For single-threaded applications, \textit{Memento+} addressed the metadata cache pollution issue well since there is no cross-thread metadata sharing and synchronization overhead.
For multi-threaded workloads, where shared metadata pollutes the cache and synchronization overhead occurs, \ASNGMalloc is more effective. \par

\noindent\textbf{\ASNGMalloc performs much better on multi-threaded workloads.}
As shown in Figure~\ref{fig_performance},  \ASNGMalloc outperforms the three state-of-the-art software allocators for all multi-threaded workloads. On average, \ASNGMalloc achieves $1.39 \times$, $1.40 \times$, $1.58 \times$, $1.73 \times$, and $1.75 \times$ speedups over Jemalloc for 1, 2, 4, 8, and 16 threads, respectively. Even when compared with the most recent software allocator TCMalloc, \ASNGMalloc achieves $1.21 \times$, $1.15 \times$, $1.18 \times$, $1.16 \times$, and $1.18 \times$ speedups for 1, 2, 4, 8, and 16 threads, respectively.  
Compared to Mimalloc, \ASNGMalloc achieves average speedups of $1.17 \times$, $1.22 \times$, $1.25 \times$, $1.21 \times$, and $1.15 \times$ for 1, 2, 4, 8, and 16 threads. Compared with two hardware accelerators \textit{Mallacc} and \textit{Memento+}, \ASNGMalloc also achieves \MaSpeedup \space and \MeSpeedup \space speedups when using 16 threads.

\noindent\textbf{Magnitude of the Butterfly Effect:}
Table~\ref{table_butter_fly} shows the percentage of allocation instructions of the overall program for the benchmarks. 
For the 10 multi-threaded workloads we used, on average \Allocation \space of the overall instructions is spent on memory allocation tasks. 
However, much like the butterfly effect, even though only \Allocation \space of total instructions is spent on memory allocation tasks, this can ultimately lead to $1.75\times$ increase in overall program performance. 
In addition, there is a large variance (from $0.4\%$ to $7.2\%$) of the overall instructions spent on memory allocation. 
Such variance indicates that the chosen workloads are covering a large workload space encompassing allocation-intensive and non-intensive workloads. 
\ASNGMalloc achieves performance improvement in all scenarios. 
\par

\begin{table}[t]
\caption{\small{Percentage of instructions on memory allocation vs overall program speedups (over Jemalloc using 16 threads). \ASNGMalloc improves the performance globally (more than just the allocator)
due to the removal of metadata cache pollution and across-core synchronization.  
}}
\label{table_butter_fly}
\centering
\setlength{\tabcolsep}{1pt}
\renewcommand{\arraystretch}{1}
\begin{tabular}{ >{\centering\arraybackslash}p{1.6cm} 
                   >{\centering\arraybackslash}p{2.6cm} 
                   >{\centering\arraybackslash}p{1.8cm} 
                   >{\centering\arraybackslash}p{1.8cm} 
                   >{\centering\arraybackslash}p{2cm} 
                   }
 \hline
\textbf{Workload} & \textbf{Allocator Ins (\%)} & \textbf{TCMalloc} & \textbf{Mimalloc} & \textbf{\ASNGMalloc} \\ 
\hline  
\textsl{Larson} & $5.99\%$ & $2.71\times$  & $2.17\times$ & $3.19\times$   \\
\textsl{Xmalloc} & $2.45\%$ & $1.06\times$  & $1.09\times$ & $1.16\times$   \\
\textsl{Scratch} & $2.62\%$ & $1.49\times$  & $1.54\times$ & $1.62\times$   \\
\textsl{Sh6bench} & $5.55\%$ & $1.63\times$  & $1.45\times$ & $1.73\times$   \\
\textsl{Sh8bench} & $7.22\%$ & $1.31\times$  & $1.39\times$ & $1.49\times$   \\
\textsl{Mstress} & $5.46\%$ & $1.65\times$  & $1.62\times$ & $1.71\times$   \\
\textsl{AllocTest} & $3.91\%$ & $1.04\times$  & $1.40\times$ & $1.46\times$   \\
\textsl{BFS} & $3.07\%$ & $2.55\times$  & $2.50\times$ & $3.57\times$   \\
\textsl{BC} & $0.37\%$ & $1.18\times$  & $1.16\times$ & $1.20\times$   \\
\textsl{DC} & $6.94\%$ & $1.10\times$  & $1.39\times$ & $1.64\times$   \\
\hline
\textbf{geo-mean} & $4.36\%$  & $1.48\times$  & $1.52\times$ & $1.75\times$   \\
\hline

\end{tabular}
\end{table}

\begin{figure}[t]
    \centerline{\includegraphics[width=0.8\columnwidth,trim = 4mm 4mm 3mm 4mm, clip=true]{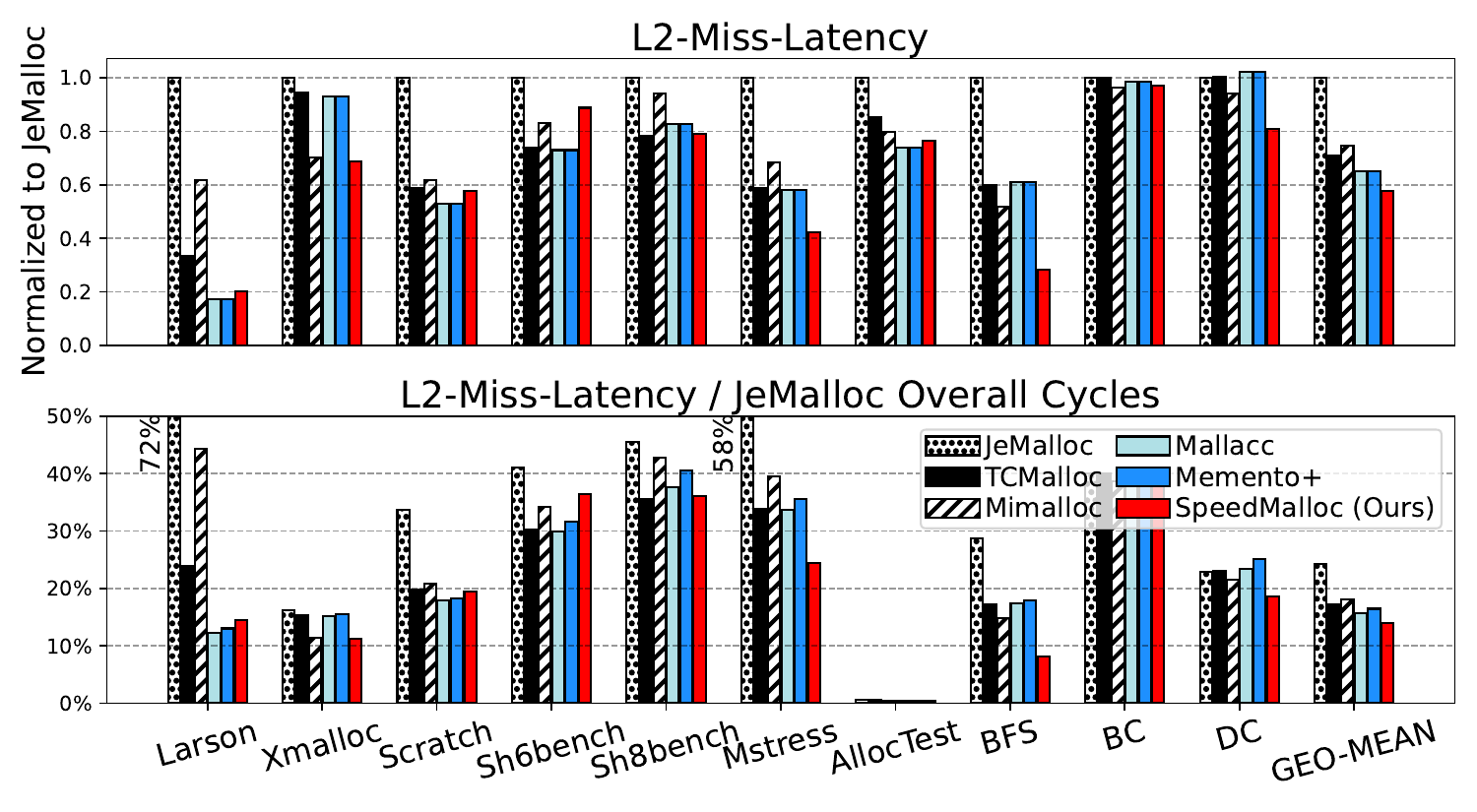}}
	\caption{\small{L2 miss cycles (of 16-thread results). \ASNGMalloc reduced 40\%+ L2 miss cycles over Jemalloc.} }
	\label{fig_cache_8}
\end{figure}

\subsubsection{Analysis} 
\label{section_result_analysis}
We analyze the sources of the performance gains. We break down the execution time to analyze the performance gains from reduced cache pollution and eliminating metadata synchronization. \par

\noindent\textbf{\ASNGMalloc reduces cache pollution.}
Figure~\ref{fig_cache_8} shows the L2 miss cycles\footnote{ We cannot get the cache line evictions due to metadata access for allocators using aggregated metadata layout (Mimalloc). Instead, we use the L2 miss cycles as a global metric to indicate reducing cache miss results in better allocator performance.} 
on the 16-core system (running 16 threads).
Since all allocator metadata are stored in the \SupportCore, \ASNGMalloc achieves $42.36\%$, $18.76\%$, and $22.80\%$ L2 miss cycles reduction compared to  Jemalloc, TCMalloc, and Mimalloc, respectively.
The L2 miss has a non-negligible impact on the program performance, which contributes $24.22\%$, $17.18\%$, and $18.08\%$ of the overall execution cycles of the 10 workloads (GEO-MEAN) when using Jemalloc, TCMalloc, and Mimalloc.
Considering the speedups for the entire program, $10.26\%$ (out of \JeSpeedupP) of the \ASNGMalloc execution time reduction over Jemalloc is due to the reduction of L2 misses. Likewise, $4.78\%$  (out of \TCSpeedupP) and $6.28\%$ (out of \MiSpeedupP) of execution time reduction over TCMalloc and Mimalloc come from L2 miss reduction.
\textit{Mallacc} and \textit{Memento+} used distributed malloc cache and accelerators for memory allocation, which did not address the shared cache metadata pollution in multi-threaded scenarios. As a result, the $3.73\%$ (out of \MeSpeedupP) performance improvement of \ASNGMalloc over \textit{Memento+} comes from the L2 miss reduction. \par

\begin{figure}[t]	\centerline{\includegraphics[width=0.9\columnwidth,trim = 4mm 0mm 3mm 3mm, clip=true]{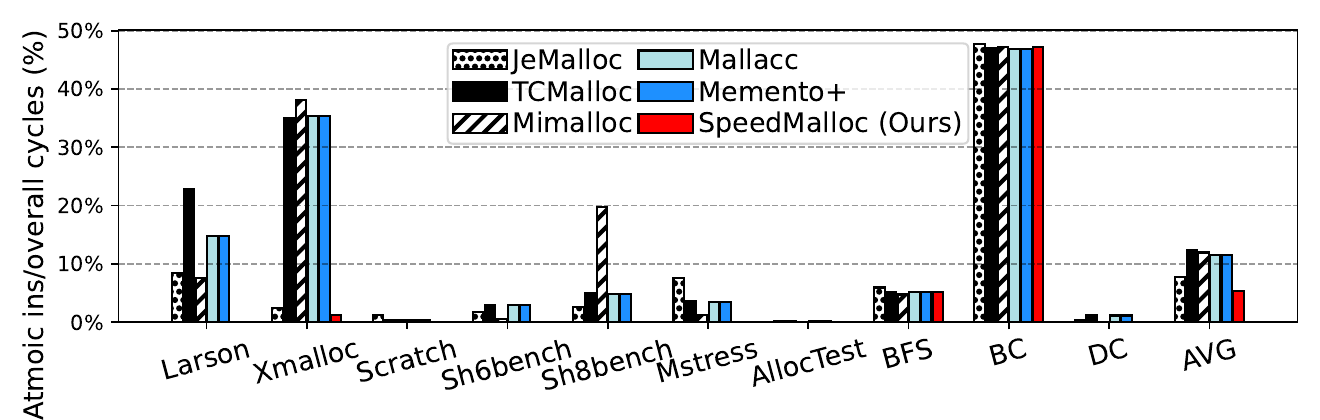}}
	\caption{\small{Atomic instruction cycles / Jemalloc overall cycles (of 16-thread results). $\sim$$10\%$ of overall time is saved due to the removal of cross-core metadata synchronization in \ASNGMalloc.} }
	\label{fig_atomic_result}
\end{figure}

\noindent\textbf{\ASNGMalloc eliminates metadata synchronization.}
\ASNGMalloc eliminates metadata synchronization incurred in multi-threaded software allocators by removing atomic instructions. 
We quantify the overhead of these atomic instructions in other allocators (shown in Figure~\ref{fig_atomic_result}). 
\ASNGMalloc reduces the atomic instruction cycles for all workloads. 
Considering the average of the 10 multi-threaded workloads, $3.57\%$ (out of \JeSpeedupP), $11.99\%$  (out of \TCSpeedupP), $11.73\%$  (out of \MiSpeedupP), $10.34\%$  (out of \MaSpeedupP), and $10.81\%$  (out of \MeSpeedupP) of the \ASNGMalloc execution time reduction over Jemalloc, TCMalloc, Mimalloc, \textit{Mallacc}, and \textit{Memento+} is due to the removal of cross-core metadata synchronization. 
\par

\noindent\textbf{Other factors affecting performance.} 
The average performance gains (of the 10 multi-threaded workloads) of \ASNGMalloc over TCMalloc and Mimalloc are almost consistent with the two individual factors 
(e.g., for Mimalloc \MiSpeedupP \space $\approx$ $6.28\% + 11.73\%$).
There are also other potential factors affecting allocator performance. \ASNGMalloc also benefits from these aspects, which results in more speedups over Jemalloc. \par

First, \ASNGMalloc achieves a reduction in the number of instructions over Jemalloc. 
We built \ASNGMalloc atop TCMalloc and Mimalloc.
On average of the 10 workloads, TCMalloc and Mimalloc saves 11.1\% and 13.9\% instructions compared with Jemalloc, respectively.
The number of instructions saved is even larger than the time spent on memory allocation. 
The reason is that TCMalloc and Mimalloc handle address mapping well and help reduce the number of page faults (which are handled out of the allocation phase in the kernel space). 
In addition, with the removal of metadata contention, \ASNGMalloc uses shared metadata (to indicate available free spaces) for all threads. 
This makes \ASNGMalloc achieve an additional 4.97\% instruction savings over TCMalloc.
Cache line false sharing is another reason, which indicates multiple small objects are allocated within the same cache line but shared by multiple cores. Mimalloc and TCMalloc handle cache line false sharing better than Jemalloc~\cite{leijen2019mimalloc} (\textsl{Scratch} exemplifies this aspect). \par

Memory allocation is sophisticated and can affect the program execution from many perspectives. In addition to performance gains from reducing cache pollution and eliminating metadata synchronization, \ASNGMalloc can also benefit from these other factors mentioned above. \par

\begin{figure}[t]
	\centerline{\includegraphics[width=0.9\columnwidth,trim = 3mm 5mm 3mm 3mm, clip=true]{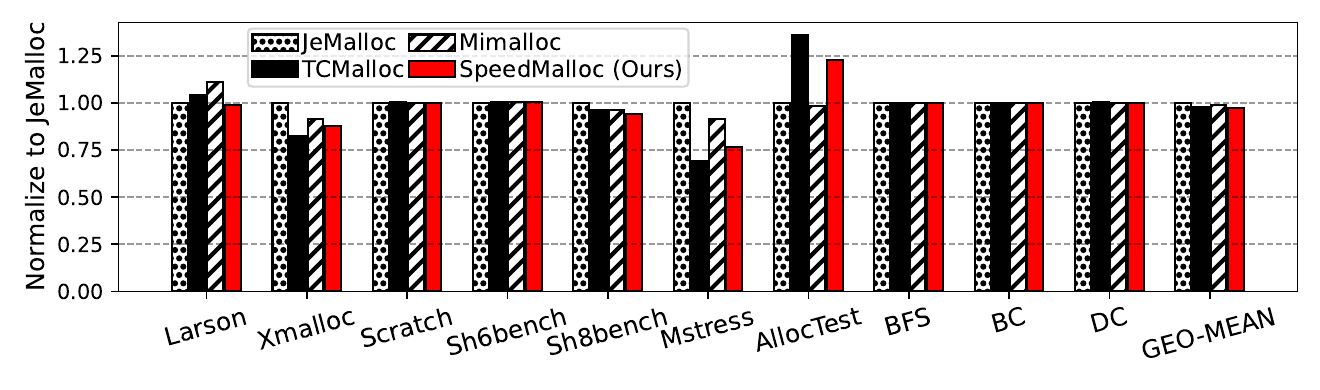}}
	\caption{\small{Memory consumption (16-thread results, smaller is better). \ASNGMalloc consumes similar memory as other allocators. }}
	\label{fig_memory_consumption}
\end{figure}

\noindent\textbf{Memory Consumption.}
We evaluate the memory consumption of \ASNGMalloc.
\ASNGMalloc adopts state-of-the-art software allocator design and addresses \textit{blowup} well.
\ASNGMalloc can achieve more memory savings due to the use of centralized metadata, which reduces memory fragmentation in thread-local data.
However, \ASNGMalloc may also trigger additional memory consumption since the \SupportCore priorities allocation requests,
delaying recycling memory from deallocation requests, which increases peak memory consumption (Figure~\ref{fig_message_queue}).
Overall (shown in Figure~\ref{fig_memory_consumption}), \ASNGMalloc consumes nearly the same ($\sim$$1\%$ difference) memory compared with TCMalloc and Mimalloc (\textit{Mallacc} and \textit{Memento+} are built atop TCMalloc, we assume they have the same memory consumption). \par

%% file: 08_03_evaluation.tex
\ASNGMalloc also achieves energy savings, although the added \SupportCore consumes additional power. 
Compared with a regular out-of-order core, the \SupportCore is a simple in-order core without any floating point or vector units (and registers), instruction renaming units, re-order buffers, etc. 
As a result, the power of a \SupportCore is estimated to be $33.72\%$ that of a main core, and the area of a \SupportCore is only $24.43\%$ that of a main core (results obtained using McPAT~\cite{li2009mcpat}). Considering a 16-core system (with memory controllers), the \SupportCore only consumes $1.50\%$ and $2.25\%$ of the entire process chip area and power, respectively. \par

\begin{figure}[t]
    \centerline{\includegraphics[width=0.7\columnwidth,trim = 2mm 1mm 3mm 3mm, clip=true]{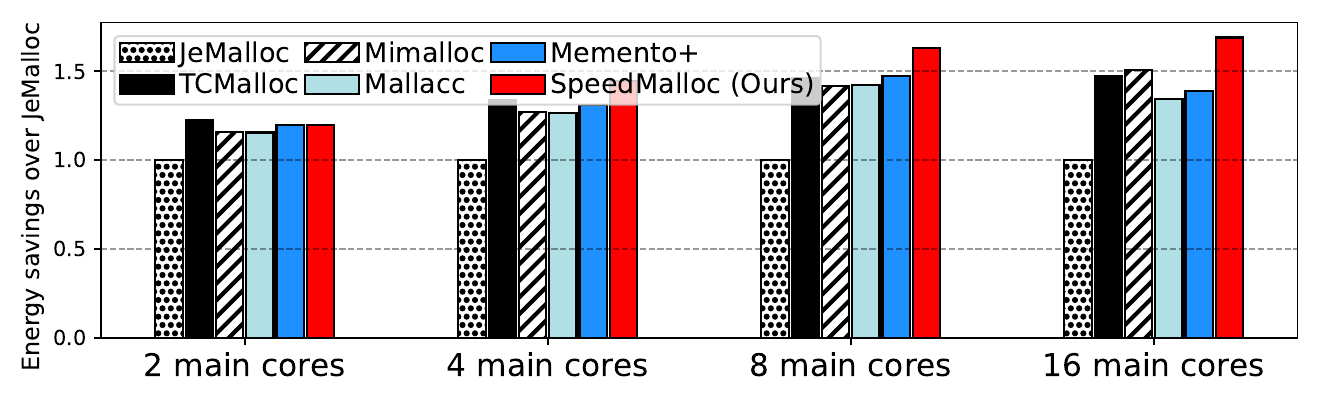}}
	\caption{\small{Energy Savings (the larger the better). \ASNGMalloc saves energy compared to other allocators. }}
	\label{fig_energy}
\end{figure}

Figure~\ref{fig_energy} shows that \ASNGMalloc achieves energy savings over the three software allocators when using 2 or more main cores. On average of the 10 multi-threaded workloads, \ASNGMalloc achieves  $1.20 \times$, $1.44 \times$, $1.63 \times$, and $1.69 \times$ energy savings over Jemalloc for 2, 4, 8, and 16 main cores; 
$0.98 \times$, $1.08 \times$, $1.11 \times$, and $1.15 \times$ compared to TCMalloc;  
$1.03 \times$, $1.13 \times$, $1.15 \times$, and $1.12 \times$ compared to Mimalloc.
Comparing against the two hardware accelerator solutions \textit{Mallacc} and \textit{Memento+}, \ASNGMalloc achieves \MaEnergy \space and \MeEnergy \space savings, as these hardware solutions require additional hardware resources for each core.
Therefore, using a \SupportCore for memory allocation is feasible since the power overhead is negligible. \par

\begin{figure}[t]
    \centerline{\includegraphics[width=0.8\columnwidth,trim = 2mm 3mm 3mm 3mm, clip=true]{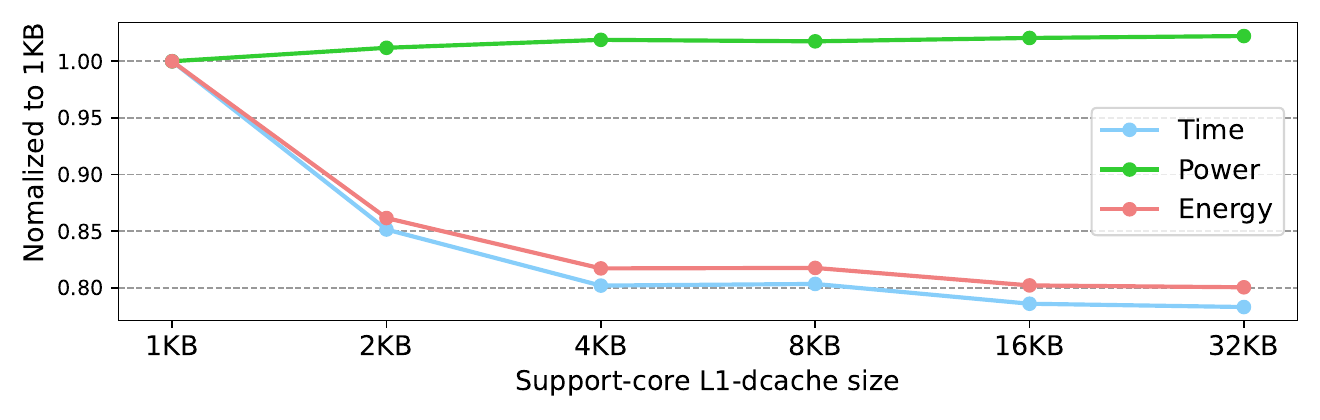}}
	\caption{\small{Sensitivity on \SupportCore L1-dcache capacity. All numbers normalized to 1KB L1-dcache (the smaller the better). 16KB is the most energy-efficient design.}}
	\label{fig_cache_size}
\end{figure}

We also perform sensitivity studies on the \SupportCore microarchitecture configures. Figure~\ref{fig_cache_size} shows the execution time, power, and energy of running workload \textsl{Sh6bench} using 16 main cores (with one \SupportCore). By increasing the L1-dcache capacity from 1KB to 16KB, there is a $1.27\times$ speedup while only $2.1\%$ additional power consumption. Overall, 16KB L1-dcache capacity is the most energy-efficient design (\ASNGMalloc also has the flexibility of using a 4KB L1-dcache in power-constrained embedded systems).
\par

%% file: 08_04_evaluation.tex
\ASNGMalloc achieves better performance than state-of-the-art software memory allocators, by introducing additional architectural support over the current system. 
Other potential solutions may not require hardware changes: using cache partitions (\S~\ref{section_cache_partition}) or harvesting an idle core in the system (\S~\ref{section_idle_core}).
However, these solutions have their limitations and cannot achieve the efficiency of \ASNGMalloc.

\subsubsection{\ASNGMalloc vs. Cache partition}
\label{section_cache_partition}
Cache pollution may be eliminated by partitioning caches and allocating dedicated ways (assuming a set-associative cache) to allocator metadata. 
However, dedicating ways for allocator metadata may adversely impact some applications since less amount of cache is available for application user data. On the other hand, \ASNGMalloc is a more workload-agnostic solution. 
We compare \ASNGMalloc with cache partition in this section. \par

\begin{figure}[t]
    \centerline{\includegraphics[width=0.9\columnwidth,trim = 4mm 5mm 3mm 4mm, clip=true]{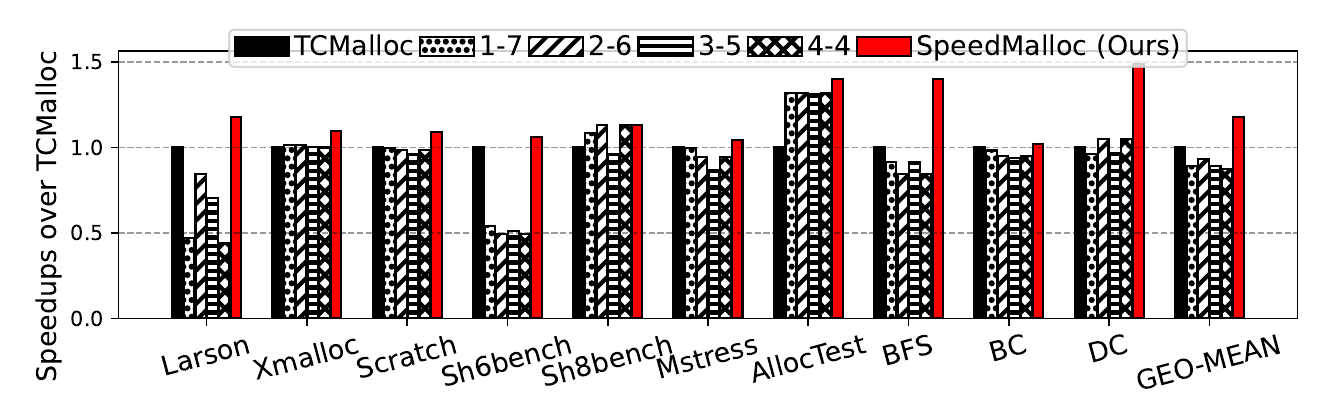}}
	\caption{\small{Employing cache partition (16 threads). Cache partition works for some cases, but \ASNGMalloc is a more general solution.}}
	\label{fig_partition}
\end{figure}

We partition the 8-way L2 into two groups, one dedicated to allocator metadata and the other for all other program data, e.g., in Figure~\ref{fig_partition}, 7-1 indicates that 1 way is exclusive for allocator metadata.
As the results suggest, the cache partition does not secure a performance gain on all applications and results in a $7\%$$\sim$$12\%$ slowdown. \par

The partition schema is workload-dependent, thus, using a static partition will not achieve performance gains for all workloads.
Dynamic cache partition has been used in prior works~\cite{sanchez2011vantage, mittal2017survey}, but still faces many challenges. 
First, the overhead of additional control logic for tracking cache usage to determine proper partition is non-negligible~\cite{sanchez2011vantage, mittal2017survey}.  
Additionally, whether a $malloc$ request invokes $mmap$ system calls is unpredictable, which results in significant memory access difference~\cite{tcmalloc, mimalloc}.  
These challenges prevent cache partition from becoming a viable alternative to \ASNGMalloc. 
\ASNGMalloc is a more general solution and improves the performance for all workloads, not only because it addresses cache pollution, but also multi-threaded metadata synchronization (\S~\ref{section_result_analysis}). \par

\begin{figure}[t]
    \centerline{\includegraphics[width=0.9\columnwidth,trim = 4mm 5mm 3mm 3.5mm, clip=true]{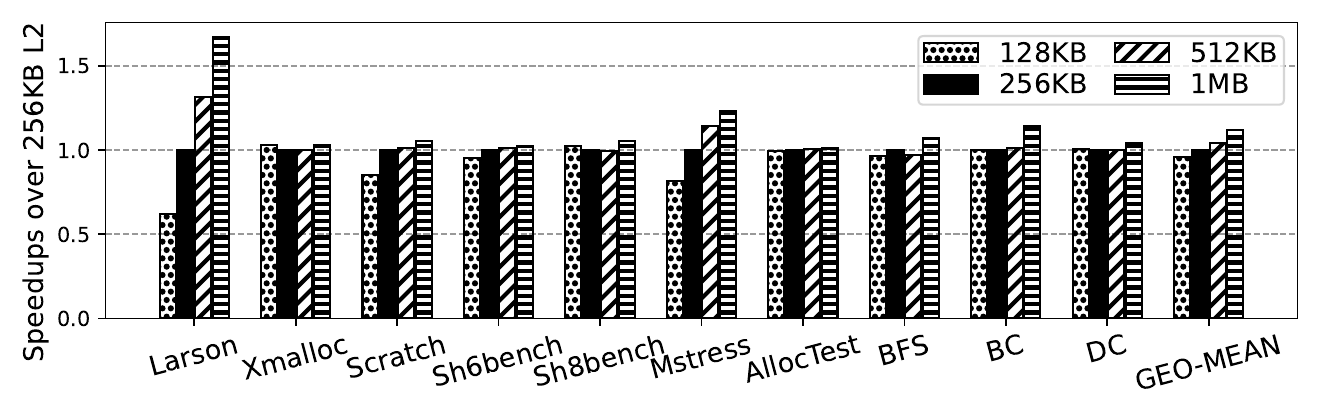}}
	\caption{\small{Sensitivity of L2 cache capacity on the 16-core system. Increasing L2 cache capacity is not a one-fit-all solution.  }}
	\label{fig_sensitivity}
\end{figure}

\noindent\textbf{Possibility of increasing cache capacity.} 
Instead of cache partition, increasing cache capacity may reduce cache pollution. 
We show the performance of Mimalloc using different cache capacities in Figure~\ref{fig_sensitivity} (256KB L2 is the baseline in Table~\ref{tab:gem5_settings}). 
On average for the 10 workloads, increasing the cache capacity by $2 \times$ (from 128KB to 256KB) can result in $1.04 \times$ speedup and increasing cache size by $8 \times$ (from 128KB to 1MB) can result in $1.17 \times$ speedup. \par

We find that increasing cache capacity can lead to performance improvements in workloads that are bounded by cache misses.
For example, the performance Mimalloc on \textsl{LarsonN} is more bounded by cache misses (Figure~\ref{fig_cache_8}) instead of atomic cycles (Figure~\ref{fig_atomic_result}), so increasing L2 cache capacity can also bring performance improvement. 
However, increasing the cache capacity for workloads that are bounded by cache misses, e.g., \textsl{AllocTest}, could be wasteful.
Moreover, increasing cache capacity (either makes the additional capacity exclusively for memory allocators or not): (a) cannot resolve the \textit{metadata synchronization} issue (\S~\ref{section_result_analysis}), (b) may help with metadata capacity miss but not with conflict miss, (c) cannot simplify allocator design (\S~\ref{section_result_analysis}), and (d) adds to cache access latency~\cite{7zip}, area, and power overheads. \par

\begin{figure}[t]
\centerline{\includegraphics[width=0.9\columnwidth,trim = 4mm 4mm 11mm 3mm, clip=true]{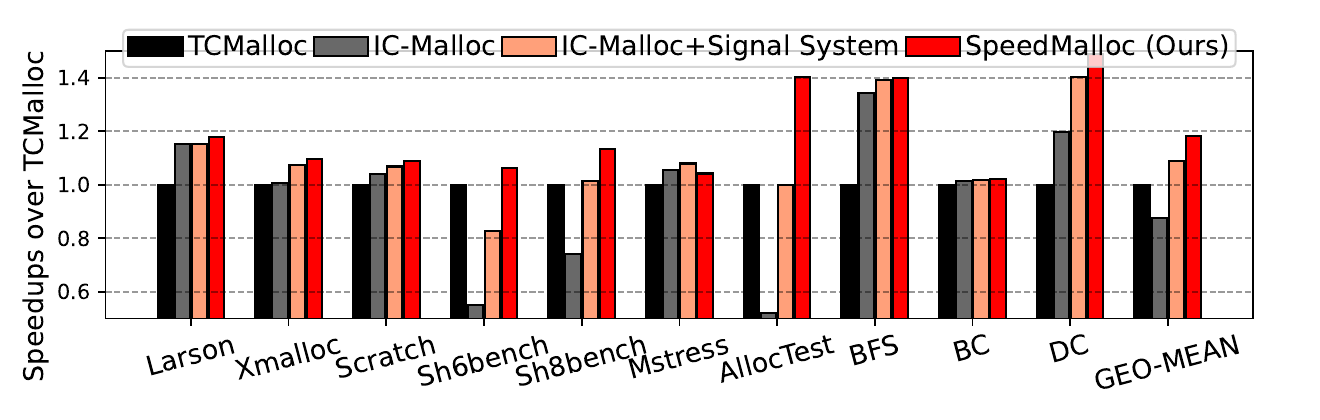}}
	\caption{\small{\ASNGMalloc achieves speedup over malloc with idle core \textit{IC-Malloc}, due to additional signal system (for fast allocation invocation) and hardware message queues (for relieving contention). }}
	\label{fig_nc_performance}
\end{figure}

\subsubsection{\ASNGMalloc vs. Harvesting an idle core}
\label{section_idle_core}
Instead of introducing an additional \SupportCore along with hardware enhancements, one may wonder if it is possible to reduce cache pollution and metadata synchronization by offloading allocation tasks to one of the idle cores. 
In this section, we compare the performance of \ASNGMalloc with using an idle core (we refer to this as \textit{IC-Malloc}) for memory allocation. We use the same \textit{gem5} setup for a fair comparison. \par

Figure~\ref{fig_nc_performance} shows the performance comparison between TCMalloc, \textit{IC-Malloc}, and \ASNGMalloc when using 16 threads. 
Although \textit{IC-Malloc} effectively reduces cache conflicts, its scalability is constrained by existing cross-core communication interfaces.
\textit{IC-Malloc} cannot outperform TCMalloc considering the average of the 10 workloads. 
In addition, using a full-fledged core for memory allocation tasks is unnecessary and leads to higher energy consumption. 
\par

\ASNGMalloc outperforms \textit{IC-Malloc} with the invocation interface and runtime support. 
The signal system (\S~\ref{section_control}) in \ASNGMalloc is necessary for a fast allocation invocation.
The hardware message queues (\S~\ref{section_hardware}) in the \SupportCore, which helps buffering and scheduling allocation requests in allocation-intensive workloads, is missing in \textit{IC-Malloc}.
Without these supports, the cross-core communication in \textit{IC-Malloc} can only be implemented using atomic instructions with the current cache coherency model. 
We break down the performance benefits of these supports in Figure~\ref{fig_nc_performance}. 
Decoupling the allocator from the main process (\S~\ref{section_software_approach}) cannot make \textit{IC-Malloc} achieve better performance than TCMalloc. 
With the signal system for fast invocation (\S~\ref{section_control}), a $1.09 \times$ speedup over TCMalloc can be achieved. 
Finally, with the hardware message queues to relieve contention (\S~\ref{section_hardware}), \ASNGMalloc can achieve a \TCSpeedup \space speedup over TCMalloc. \par

%% file: 08_06_evaluation.tex
Our analysis (\S~\ref{section_performance}) shows that one \SupportCore can serve memory allocation requests from 16 main cores. In modern many-core workstation systems, e.g., Apple's M3 Max chip which has 12 performance cores and 4 efficiency cores. We suggest using one of the efficiency cores as the \textit{support} core. We also discuss how \ASNGMalloc scales for larger warehouse-scale computers in this section. \par

To scale for warehouse-scale computers (64+ cores), \ASNGMalloc can employ multiple \textit{support} cores.
For example, contemporary systems such as AMD's Zen architecture~\cite{naffziger2021pioneering, bhargava2024amd} have eight cores per chiplet and Intel's recent Xeon architecture has 15 cores per chiplet~\cite{nassif2022sapphire}; one can consider adding one \SupportCore per core complex.
This setup is suitable for container-based datacenter hyperscale workloads~\cite{stojkovic2023mumanycore, sahraei2023xfaas}, where cross-chiplet data transfer is minimal.
For monolithic architecture, \ASNGMalloc could connect all \textit{support-cores} in a manner similar to the connection of last-level cache slices in scalable coherency fabric, such as in the Grace super chip~\cite{elster2022nvidia, grace}.
\ASNGMalloc could also use a more powerful \SupportCore for memory allocation-intensive workloads. 
This remains an energy-efficient solution, as the power budget of the \SupportCore increases with the number of main cores. 
Additionally, \ASNGMalloc could optimize the allocator software design by using a local cache for every 16 cores to reduce cross-\textit{support-core} metadata synchronization. \par

%% file: 010_conclusion.tex
Memory allocator performance is an important issue as demonstrated by many recent works~\cite{hunter2021beyond, zhou2024characterizing}. 
Many efforts have been reported to improve memory allocators over the past few decades, but few have addressed the interference from the allocator metadata in multi-threaded applications. \par

In this paper, we presented \ASNGMalloc by using a lightweight \SupportCore for memory allocation tasks. 
With \ASNGMalloc the overall program performance benefits from reducing cache pollution by not bringing metadata into main core caches and removing unnecessary inter-core metadata synchronization. 
We compared \ASNGMalloc with Meta's Jemalloc, Google's TCMalloc, Microsoft's Mimalloc, and two accelerators \textit{Mallacc} and \textit{Memento+}.
With a full system simulation of the entire program, we have shown that \ASNGMalloc achieves \JeSpeedup \space speedup and \JeEnergy \space energy savings over Jemalloc.
When compared to TCMalloc and  Mimalloc, \ASNGMalloc achieves \TCSpeedup \space and \MiSpeedup \space speedups, and \TCEnergy \space and \MiEnergy \space energy savings.
Compared with accelerators, \ASNGMalloc also achieves \MaSpeedup \space and \MeSpeedup \space speedups over state-of-the-art accelerators \textit{Mallacc} and \textit{Memento+}.  \par

The minimal hardware modifications in the \SupportCore make it programmable and capable of offloading additional system functions. Since memory allocation is a fundamental component in all applications, the cumulative effect of multiple optimizations, each improving performance by a few percent, can result in substantial CPU cycle savings, particularly when deployed at scale across megawatt datacenters. \par

%% file: 00_abstract.bbl

\begin{thebibliography}{61}


\ifx \showCODEN    \undefined \def \showCODEN     #1{\unskip}     \fi
\ifx \showISBNx    \undefined \def \showISBNx     #1{\unskip}     \fi
\ifx \showISBNxiii \undefined \def \showISBNxiii  #1{\unskip}     \fi
\ifx \showISSN     \undefined \def \showISSN      #1{\unskip}     \fi
\ifx \showLCCN     \undefined \def \showLCCN      #1{\unskip}     \fi
\ifx \shownote     \undefined \def \shownote      #1{#1}          \fi
\ifx \showarticletitle \undefined \def \showarticletitle #1{#1}   \fi
\ifx \showURL      \undefined \def \showURL       {\relax}        \fi
\providecommand\bibfield[2]{#2}
\providecommand\bibinfo[2]{#2}
\providecommand\natexlab[1]{#1}
\providecommand\showeprint[2][]{arXiv:#2}

\bibitem[7zi(2024)]%
        {7zip}
 \bibinfo{year}{2024}\natexlab{}.
\newblock \bibinfo{title}{7-Zip LZMA Benchmark}.
\newblock
  \bibinfo{howpublished}{\url{https://www.7-cpu.com/cpu/Skylake_X.html/}}.
\newblock


\bibitem[gem(2024)]%
        {gem5cpu}
 \bibinfo{year}{2024}\natexlab{}.
\newblock \bibinfo{title}{gem5 CPU Models}.
\newblock
  \bibinfo{howpublished}{\url{https://www.gem5.org/documentation/general_docs/cpu_models/}}.
\newblock


\bibitem[red(2024)]%
        {redis-benchmark}
 \bibinfo{year}{2024}\natexlab{}.
\newblock \bibinfo{title}{Redis-benchmark}.
\newblock
  \bibinfo{howpublished}{\url{https://redis.io/docs/management/optimization/benchmarks/}}.
\newblock


\bibitem[AG(2018)]%
        {alloc-test}
\bibfield{author}{\bibinfo{person}{Technologies AG}.}
  \bibinfo{year}{2018}\natexlab{}.
\newblock \bibinfo{title}{alloc-test}.
\newblock
  \bibinfo{howpublished}{\url{http://ithare.com/testing-memory-allocators-ptmalloc2-tcmalloc-hoard-jemalloc-while-trying-to-simulate-real-world-loads/}}.
\newblock


\bibitem[Aigner et~al\mbox{.}(2015)]%
        {aigner2015fast}
\bibfield{author}{\bibinfo{person}{Martin Aigner}, \bibinfo{person}{Christoph~M
  Kirsch}, \bibinfo{person}{Michael Lippautz}, {and} \bibinfo{person}{Ana
  Sokolova}.} \bibinfo{year}{2015}\natexlab{}.
\newblock \showarticletitle{Fast, multicore-scalable, low-fragmentation memory
  allocation through large virtual memory and global data structures}.
\newblock \bibinfo{journal}{\emph{ACM SIGPLAN Notices}} \bibinfo{volume}{50},
  \bibinfo{number}{10} (\bibinfo{year}{2015}), \bibinfo{pages}{451--469}.
\newblock


\bibitem[Asgharzadeh et~al\mbox{.}(2022)]%
        {asgharzadeh2022free}
\bibfield{author}{\bibinfo{person}{Ashkan Asgharzadeh}, \bibinfo{person}{Juan~M
  Cebrian}, \bibinfo{person}{Arthur Perais}, \bibinfo{person}{Stefanos
  Kaxiras}, {and} \bibinfo{person}{Alberto Ros}.}
  \bibinfo{year}{2022}\natexlab{}.
\newblock \showarticletitle{Free atomics: hardware atomic operations without
  fences.}. In \bibinfo{booktitle}{\emph{ISCA}}. \bibinfo{pages}{14--26}.
\newblock


\bibitem[Bailey et~al\mbox{.}(1991)]%
        {bailey1991parallel}
\bibfield{author}{\bibinfo{person}{David~H Bailey}, \bibinfo{person}{Eric
  Barszcz}, \bibinfo{person}{John~T Barton}, \bibinfo{person}{David~S
  Browning}, \bibinfo{person}{Robert~L Carter}, \bibinfo{person}{Leonardo
  Dagum}, \bibinfo{person}{Rod~A Fatoohi}, \bibinfo{person}{Paul~O
  Frederickson}, \bibinfo{person}{Thomas~A Lasinski}, \bibinfo{person}{Rob~S
  Schreiber}, {et~al\mbox{.}}} \bibinfo{year}{1991}\natexlab{}.
\newblock \showarticletitle{The NAS parallel benchmarks—summary and
  preliminary results}. In \bibinfo{booktitle}{\emph{Proceedings of the 1991
  ACM/IEEE Conference on Supercomputing}}. \bibinfo{pages}{158--165}.
\newblock


\bibitem[Barrett(2024)]%
        {cfrac}
\bibfield{author}{\bibinfo{person}{Dave Barrett}.}
  \bibinfo{year}{2024}\natexlab{}.
\newblock \bibinfo{title}{Mstress}.
\newblock
  \bibinfo{howpublished}{\url{https://github.com/microsoft/test-suite/tree/master/MultiSource/Benchmarks/MallocBench/cfrac/}}.
\newblock


\bibitem[Beamer et~al\mbox{.}(2015)]%
        {beamer2015gap}
\bibfield{author}{\bibinfo{person}{Scott Beamer}, \bibinfo{person}{Krste
  Asanovi{\'c}}, {and} \bibinfo{person}{David Patterson}.}
  \bibinfo{year}{2015}\natexlab{}.
\newblock \showarticletitle{The GAP benchmark suite}.
\newblock \bibinfo{journal}{\emph{arXiv preprint arXiv:1508.03619}}
  (\bibinfo{year}{2015}).
\newblock


\bibitem[Berger et~al\mbox{.}(2000)]%
        {berger2000hoard}
\bibfield{author}{\bibinfo{person}{Emery~D Berger}, \bibinfo{person}{Kathryn~S
  McKinley}, \bibinfo{person}{Robert~D Blumofe}, {and} \bibinfo{person}{Paul~R
  Wilson}.} \bibinfo{year}{2000}\natexlab{}.
\newblock \showarticletitle{Hoard: A scalable memory allocator for
  multithreaded applications}.
\newblock \bibinfo{journal}{\emph{ACM Sigplan Notices}} \bibinfo{volume}{35},
  \bibinfo{number}{11} (\bibinfo{year}{2000}), \bibinfo{pages}{117--128}.
\newblock
\href{https://doi.org/10.1145/378995.379232}{doi:\nolinkurl{10.1145/378995.379232}}


\bibitem[Berger et~al\mbox{.}(2002)]%
        {berger2002reconsidering}
\bibfield{author}{\bibinfo{person}{Emery~D Berger}, \bibinfo{person}{Benjamin~G
  Zorn}, {and} \bibinfo{person}{Kathryn~S McKinley}.}
  \bibinfo{year}{2002}\natexlab{}.
\newblock \showarticletitle{Reconsidering custom memory allocation}. In
  \bibinfo{booktitle}{\emph{Proceedings of the 17th ACM SIGPLAN conference on
  Object-oriented programming, systems, languages, and applications}}.
  \bibinfo{pages}{1--12}.
\newblock


\bibitem[Bhargava and Troester(2024)]%
        {bhargava2024amd}
\bibfield{author}{\bibinfo{person}{Ravi Bhargava} {and} \bibinfo{person}{Kai
  Troester}.} \bibinfo{year}{2024}\natexlab{}.
\newblock \showarticletitle{AMD Next Generation" Zen 4" Core and 4 th Gen AMD
  EPYC™ Server CPUs}.
\newblock \bibinfo{journal}{\emph{IEEE Micro}} (\bibinfo{year}{2024}).
\newblock


\bibitem[Binkert et~al\mbox{.}(2011)]%
        {binkert2011gem5}
\bibfield{author}{\bibinfo{person}{Nathan Binkert}, \bibinfo{person}{Bradford
  Beckmann}, \bibinfo{person}{Gabriel Black}, \bibinfo{person}{Steven~K
  Reinhardt}, \bibinfo{person}{Ali Saidi}, \bibinfo{person}{Arkaprava Basu},
  \bibinfo{person}{Joel Hestness}, \bibinfo{person}{Derek~R Hower},
  \bibinfo{person}{Tushar Krishna}, \bibinfo{person}{Somayeh Sardashti},
  {et~al\mbox{.}}} \bibinfo{year}{2011}\natexlab{}.
\newblock \showarticletitle{The gem5 simulator}.
\newblock \bibinfo{journal}{\emph{ACM SIGARCH computer architecture news}}
  \bibinfo{volume}{39}, \bibinfo{number}{2} (\bibinfo{year}{2011}),
  \bibinfo{pages}{1--7}.
\newblock


\bibitem[Boreham(2000)]%
        {boreham2000malloc}
\bibfield{author}{\bibinfo{person}{David Boreham}.}
  \bibinfo{year}{2000}\natexlab{}.
\newblock \showarticletitle{Malloc () performance in a multithreaded Linux
  environment}. In \bibinfo{booktitle}{\emph{2000 USENIX Annual Technical
  Conference (USENIX ATC 00)}}.
\newblock


\bibitem[Dang et~al\mbox{.}(2022)]%
        {dang2022nvalloc}
\bibfield{author}{\bibinfo{person}{Zheng Dang}, \bibinfo{person}{Shuibing He},
  \bibinfo{person}{Peiyi Hong}, \bibinfo{person}{Zhenxin Li},
  \bibinfo{person}{Xuechen Zhang}, \bibinfo{person}{Xian-He Sun}, {and}
  \bibinfo{person}{Gang Chen}.} \bibinfo{year}{2022}\natexlab{}.
\newblock \showarticletitle{NVAlloc: rethinking heap metadata management in
  persistent memory allocators}. In \bibinfo{booktitle}{\emph{Proceedings of
  the 27th ACM International Conference on Architectural Support for
  Programming Languages and Operating Systems}}. \bibinfo{pages}{115--127}.
\newblock


\bibitem[Elster and Haugdahl(2022)]%
        {elster2022nvidia}
\bibfield{author}{\bibinfo{person}{Anne~C Elster} {and} \bibinfo{person}{Tor~A
  Haugdahl}.} \bibinfo{year}{2022}\natexlab{}.
\newblock \showarticletitle{Nvidia hopper gpu and grace cpu highlights}.
\newblock \bibinfo{journal}{\emph{Computing in Science \& Engineering}}
  \bibinfo{volume}{24}, \bibinfo{number}{2} (\bibinfo{year}{2022}),
  \bibinfo{pages}{95--100}.
\newblock


\bibitem[Evans(2006)]%
        {evans2006scalable}
\bibfield{author}{\bibinfo{person}{Jason Evans}.}
  \bibinfo{year}{2006}\natexlab{}.
\newblock \showarticletitle{A scalable concurrent malloc (3) implementation for
  FreeBSD}. In \bibinfo{booktitle}{\emph{Proc. of the bsdcan conference,
  ottawa, canada}}.
\newblock


\bibitem[Frumkin and Shabano(2003)]%
        {frumkin2003arithmetic}
\bibfield{author}{\bibinfo{person}{Michael~A Frumkin} {and}
  \bibinfo{person}{Leonid Shabano}.} \bibinfo{year}{2003}\natexlab{}.
\newblock \showarticletitle{Arithmetic data cube as a data intensive
  benchmark}.
\newblock  (\bibinfo{year}{2003}).
\newblock


\bibitem[Gloger(2024)]%
        {ptmalloc}
\bibfield{author}{\bibinfo{person}{Wolfram Gloger}.}
  \bibinfo{year}{2024}\natexlab{}.
\newblock \bibinfo{title}{``Wolfram Gloger's malloc homepage''}.
\newblock \bibinfo{howpublished}{\url{http://www.malloc.de/en/}}.
\newblock


\bibitem[Gonzalez et~al\mbox{.}(2023)]%
        {gonzalez2023profiling}
\bibfield{author}{\bibinfo{person}{Abraham Gonzalez}, \bibinfo{person}{Aasheesh
  Kolli}, \bibinfo{person}{Samira Khan}, \bibinfo{person}{Sihang Liu},
  \bibinfo{person}{Vidushi Dadu}, \bibinfo{person}{Sagar Karandikar},
  \bibinfo{person}{Jichuan Chang}, \bibinfo{person}{Krste Asanovic}, {and}
  \bibinfo{person}{Parthasarathy Ranganathan}.}
  \bibinfo{year}{2023}\natexlab{}.
\newblock \showarticletitle{Profiling hyperscale big data processing}. In
  \bibinfo{booktitle}{\emph{Proceedings of the 50th Annual International
  Symposium on Computer Architecture}}. \bibinfo{pages}{1--16}.
\newblock


\bibitem[Google(2024)]%
        {tcmalloc}
\bibfield{author}{\bibinfo{person}{Google}.} \bibinfo{year}{2024}\natexlab{}.
\newblock \bibinfo{title}{TCMalloc}.
\newblock \bibinfo{howpublished}{\url{https://github.com/google/tcmalloc/}}.
\newblock


\bibitem[Grunwald et~al\mbox{.}(1993)]%
        {grunwald1993improving}
\bibfield{author}{\bibinfo{person}{Dirk Grunwald}, \bibinfo{person}{Benjamin
  Zorn}, {and} \bibinfo{person}{Robert Henderson}.}
  \bibinfo{year}{1993}\natexlab{}.
\newblock \showarticletitle{Improving the cache locality of memory allocation}.
  In \bibinfo{booktitle}{\emph{Proceedings of the ACM SIGPLAN 1993 conference
  on Programming language design and implementation}}.
  \bibinfo{pages}{177--186}.
\newblock


\bibitem[Hunter et~al\mbox{.}(2021)]%
        {hunter2021beyond}
\bibfield{author}{\bibinfo{person}{A.H. Hunter}, \bibinfo{person}{Chris
  Kennelly}, \bibinfo{person}{Paul Turner}, \bibinfo{person}{Darryl Gove},
  \bibinfo{person}{Tipp Moseley}, {and} \bibinfo{person}{Parthasarathy
  Ranganathan}.} \bibinfo{year}{2021}\natexlab{}.
\newblock \showarticletitle{Beyond malloc efficiency to fleet efficiency: a
  hugepage-aware memory allocator}. In \bibinfo{booktitle}{\emph{15th {USENIX}
  Symposium on Operating Systems Design and Implementation ({OSDI} 21)}}.
  \bibinfo{publisher}{{USENIX} Association}, \bibinfo{pages}{257--273}.
\newblock
\showISBNx{978-1-939133-22-9}
\urldef\tempurl%
\url{https://www.usenix.org/conference/osdi21/presentation/hunter}
\showURL{%
\tempurl}


\bibitem[Kanev et~al\mbox{.}(2015)]%
        {kanev2015profiling}
\bibfield{author}{\bibinfo{person}{Svilen Kanev}, \bibinfo{person}{Juan~Pablo
  Darago}, \bibinfo{person}{Kim Hazelwood}, \bibinfo{person}{Parthasarathy
  Ranganathan}, \bibinfo{person}{Tipp Moseley}, \bibinfo{person}{Gu-Yeon Wei},
  {and} \bibinfo{person}{David Brooks}.} \bibinfo{year}{2015}\natexlab{}.
\newblock \showarticletitle{Profiling a warehouse-scale computer}. In
  \bibinfo{booktitle}{\emph{2015 ACM/IEEE 42nd Annual International Symposium
  on Computer Architecture (ISCA)}}. \bibinfo{pages}{158--169}.
\newblock
\href{https://doi.org/10.1145/2749469.2750392}{doi:\nolinkurl{10.1145/2749469.2750392}}


\bibitem[Kanev et~al\mbox{.}(2017)]%
        {kanev2017mallacc}
\bibfield{author}{\bibinfo{person}{Svilen Kanev}, \bibinfo{person}{Sam~Likun
  Xi}, \bibinfo{person}{Gu-Yeon Wei}, {and} \bibinfo{person}{David Brooks}.}
  \bibinfo{year}{2017}\natexlab{}.
\newblock \showarticletitle{Mallacc: Accelerating Memory Allocation}. In
  \bibinfo{booktitle}{\emph{Proceedings of the Twenty-Second International
  Conference on Architectural Support for Programming Languages and Operating
  Systems}} (Xi'an, China) \emph{(\bibinfo{series}{ASPLOS '17})}.
  \bibinfo{publisher}{Association for Computing Machinery},
  \bibinfo{address}{New York, NY, USA}, \bibinfo{pages}{33–45}.
\newblock
\showISBNx{9781450344654}
\href{https://doi.org/10.1145/3037697.3037736}{doi:\nolinkurl{10.1145/3037697.3037736}}


\bibitem[Karandikar et~al\mbox{.}(2021)]%
        {karandikar2021hardware}
\bibfield{author}{\bibinfo{person}{Sagar Karandikar}, \bibinfo{person}{Chris
  Leary}, \bibinfo{person}{Chris Kennelly}, \bibinfo{person}{Jerry Zhao},
  \bibinfo{person}{Dinesh Parimi}, \bibinfo{person}{Borivoje Nikolic},
  \bibinfo{person}{Krste Asanovic}, {and} \bibinfo{person}{Parthasarathy
  Ranganathan}.} \bibinfo{year}{2021}\natexlab{}.
\newblock \showarticletitle{A hardware accelerator for protocol buffers}. In
  \bibinfo{booktitle}{\emph{MICRO-54: 54th Annual IEEE/ACM International
  Symposium on Microarchitecture}}. \bibinfo{pages}{462--478}.
\newblock


\bibitem[Karandikar et~al\mbox{.}(2023)]%
        {karandikar2023cdpu}
\bibfield{author}{\bibinfo{person}{Sagar Karandikar},
  \bibinfo{person}{Aniruddha~N Udipi}, \bibinfo{person}{Junsun Choi},
  \bibinfo{person}{Joonho Whangbo}, \bibinfo{person}{Jerry Zhao},
  \bibinfo{person}{Svilen Kanev}, \bibinfo{person}{Edwin Lim},
  \bibinfo{person}{Jyrki Alakuijala}, \bibinfo{person}{Vrishab Madduri},
  \bibinfo{person}{Yakun~Sophia Shao}, {et~al\mbox{.}}}
  \bibinfo{year}{2023}\natexlab{}.
\newblock \showarticletitle{CDPU: Co-designing Compression and Decompression
  Processing Units for Hyperscale Systems}. In
  \bibinfo{booktitle}{\emph{Proceedings of the 50th Annual International
  Symposium on Computer Architecture}}. \bibinfo{pages}{1--17}.
\newblock


\bibitem[Larson and Krishnan(1998)]%
        {larson1998memory}
\bibfield{author}{\bibinfo{person}{Per-{\AA}ke Larson} {and}
  \bibinfo{person}{Murali Krishnan}.} \bibinfo{year}{1998}\natexlab{}.
\newblock \showarticletitle{Memory allocation for long-running server
  applications}.
\newblock \bibinfo{journal}{\emph{ACM SIGPLAN Notices}} \bibinfo{volume}{34},
  \bibinfo{number}{3} (\bibinfo{year}{1998}), \bibinfo{pages}{176--185}.
\newblock


\bibitem[Leijen(2024)]%
        {mstress}
\bibfield{author}{\bibinfo{person}{Daan Leijen}.}
  \bibinfo{year}{2024}\natexlab{}.
\newblock \bibinfo{title}{Mstress}.
\newblock
  \bibinfo{howpublished}{\url{https://github.com/daanx/mimalloc-bench/tree/master/bench/mstress/}}.
\newblock


\bibitem[Leijen et~al\mbox{.}(2019)]%
        {leijen2019mimalloc}
\bibfield{author}{\bibinfo{person}{Daan Leijen}, \bibinfo{person}{Ben Zorn},
  {and} \bibinfo{person}{Leonardo de Moura}.} \bibinfo{year}{2019}\natexlab{}.
\newblock \bibinfo{booktitle}{\emph{Mimalloc: Free List Sharding in Action}}.
\newblock \bibinfo{type}{{T}echnical {R}eport} MSR-TR-2019-18.
  \bibinfo{institution}{Microsoft}.
\newblock
\urldef\tempurl%
\url{https://www.microsoft.com/en-us/research/publication/mimalloc-free-list-sharding-in-action/}
\showURL{%
\tempurl}


\bibitem[Li et~al\mbox{.}(2023)]%
        {li2023nextgen}
\bibfield{author}{\bibinfo{person}{Ruihao Li}, \bibinfo{person}{Qinzhe Wu},
  \bibinfo{person}{Krishna Kavi}, \bibinfo{person}{Gayatri Mehta},
  \bibinfo{person}{Neeraja~J Yadwadkar}, {and} \bibinfo{person}{Lizy~K John}.}
  \bibinfo{year}{2023}\natexlab{}.
\newblock \showarticletitle{NextGen-Malloc: Giving Memory Allocator Its Own
  Room in the House}. In \bibinfo{booktitle}{\emph{Proceedings of the 19th
  Workshop on Hot Topics in Operating Systems}}. \bibinfo{pages}{135--142}.
\newblock


\bibitem[Li et~al\mbox{.}(2009)]%
        {li2009mcpat}
\bibfield{author}{\bibinfo{person}{Sheng Li}, \bibinfo{person}{Jung~Ho Ahn},
  \bibinfo{person}{Richard~D Strong}, \bibinfo{person}{Jay~B Brockman},
  \bibinfo{person}{Dean~M Tullsen}, {and} \bibinfo{person}{Norman~P Jouppi}.}
  \bibinfo{year}{2009}\natexlab{}.
\newblock \showarticletitle{McPAT: An integrated power, area, and timing
  modeling framework for multicore and manycore architectures}. In
  \bibinfo{booktitle}{\emph{Proceedings of the 42nd annual ieee/acm
  international symposium on microarchitecture}}. \bibinfo{pages}{469--480}.
\newblock


\bibitem[Li{\'e}tar et~al\mbox{.}(2019)]%
        {lietar2019snmalloc}
\bibfield{author}{\bibinfo{person}{Paul Li{\'e}tar}, \bibinfo{person}{Theodore
  Butler}, \bibinfo{person}{Sylvan Clebsch}, \bibinfo{person}{Sophia
  Drossopoulou}, \bibinfo{person}{Juliana Franco}, \bibinfo{person}{Matthew~J
  Parkinson}, \bibinfo{person}{Alex Shamis}, \bibinfo{person}{Christoph~M
  Wintersteiger}, {and} \bibinfo{person}{David Chisnall}.}
  \bibinfo{year}{2019}\natexlab{}.
\newblock \showarticletitle{Snmalloc: a message passing allocator}. In
  \bibinfo{booktitle}{\emph{Proceedings of the 2019 ACM SIGPLAN International
  Symposium on Memory Management}}. \bibinfo{pages}{122--135}.
\newblock


\bibitem[Lorenz(1963)]%
        {lorenz1963deterministic}
\bibfield{author}{\bibinfo{person}{Edward~N Lorenz}.}
  \bibinfo{year}{1963}\natexlab{}.
\newblock \showarticletitle{Deterministic nonperiodic flow}.
\newblock \bibinfo{journal}{\emph{Journal of atmospheric sciences}}
  \bibinfo{volume}{20}, \bibinfo{number}{2} (\bibinfo{year}{1963}),
  \bibinfo{pages}{130--141}.
\newblock


\bibitem[Lowe-Power et~al\mbox{.}(2020)]%
        {lowe2020gem5}
\bibfield{author}{\bibinfo{person}{Jason Lowe-Power},
  \bibinfo{person}{Abdul~Mutaal Ahmad}, \bibinfo{person}{Ayaz Akram},
  \bibinfo{person}{Mohammad Alian}, \bibinfo{person}{Rico Amslinger},
  \bibinfo{person}{Matteo Andreozzi}, \bibinfo{person}{Adri{\`a} Armejach},
  \bibinfo{person}{Nils Asmussen}, \bibinfo{person}{Brad Beckmann},
  \bibinfo{person}{Srikant Bharadwaj}, {et~al\mbox{.}}}
  \bibinfo{year}{2020}\natexlab{}.
\newblock \showarticletitle{The gem5 simulator: Version 20.0+}.
\newblock \bibinfo{journal}{\emph{arXiv preprint arXiv:2007.03152}}
  (\bibinfo{year}{2020}).
\newblock


\bibitem[Maas et~al\mbox{.}(2020)]%
        {maas2020asplos}
\bibfield{author}{\bibinfo{person}{Martin Maas}, \bibinfo{person}{David~G.
  Andersen}, \bibinfo{person}{Michael Isard}, \bibinfo{person}{Mohammad~Mahdi
  Javanmard}, \bibinfo{person}{Kathryn~S. McKinley}, {and}
  \bibinfo{person}{Colin Raffel}.} \bibinfo{year}{2020}\natexlab{}.
\newblock \showarticletitle{Learning-Based Memory Allocation for C++ Server
  Workloads}. In \bibinfo{booktitle}{\emph{Proceedings of the Twenty-Fifth
  International Conference on Architectural Support for Programming Languages
  and Operating Systems}} (Lausanne, Switzerland)
  \emph{(\bibinfo{series}{ASPLOS '20})}. \bibinfo{publisher}{Association for
  Computing Machinery}, \bibinfo{address}{New York, NY, USA},
  \bibinfo{pages}{541–556}.
\newblock
\showISBNx{9781450371025}
\href{https://doi.org/10.1145/3373376.3378525}{doi:\nolinkurl{10.1145/3373376.3378525}}


\bibitem[Maas et~al\mbox{.}(2021)]%
        {maas2021adaptive}
\bibfield{author}{\bibinfo{person}{Martin Maas}, \bibinfo{person}{Chris
  Kennelly}, \bibinfo{person}{Khanh Nguyen}, \bibinfo{person}{Darryl Gove},
  \bibinfo{person}{Kathryn~S. McKinley}, {and} \bibinfo{person}{Paul Turner}.}
  \bibinfo{year}{2021}\natexlab{}.
\newblock \showarticletitle{Adaptive Huge-Page Subrelease for Non-Moving Memory
  Allocators in Warehouse-Scale Computers}. In
  \bibinfo{booktitle}{\emph{Proceedings of the 2021 ACM SIGPLAN International
  Symposium on Memory Management}} (Virtual, Canada)
  \emph{(\bibinfo{series}{ISMM 2021})}. \bibinfo{publisher}{Association for
  Computing Machinery}, \bibinfo{address}{New York, NY, USA},
  \bibinfo{pages}{28–38}.
\newblock
\showISBNx{9781450384483}
\href{https://doi.org/10.1145/3459898.3463905}{doi:\nolinkurl{10.1145/3459898.3463905}}


\bibitem[Margaritov et~al\mbox{.}(2021)]%
        {margaritov2021ptemagnet}
\bibfield{author}{\bibinfo{person}{Artemiy Margaritov},
  \bibinfo{person}{Dmitrii Ustiugov}, \bibinfo{person}{Amna Shahab}, {and}
  \bibinfo{person}{Boris Grot}.} \bibinfo{year}{2021}\natexlab{}.
\newblock \showarticletitle{Ptemagnet: Fine-grained physical memory reservation
  for faster page walks in public clouds}. In
  \bibinfo{booktitle}{\emph{Proceedings of the 26th ACM International
  Conference on Architectural Support for Programming Languages and Operating
  Systems}}. \bibinfo{pages}{211--223}.
\newblock


\bibitem[MicroQuill(2024)]%
        {shbench}
\bibfield{author}{\bibinfo{person}{MicroQuill}.}
  \bibinfo{year}{2024}\natexlab{}.
\newblock \bibinfo{title}{shbench}.
\newblock \bibinfo{howpublished}{\url{http://www.microquill.com/}}.
\newblock


\bibitem[Microsoft(2024)]%
        {mimalloc-bench}
\bibfield{author}{\bibinfo{person}{Microsoft}.}
  \bibinfo{year}{2024}\natexlab{}.
\newblock \bibinfo{title}{Mimalloc-bench}.
\newblock
  \bibinfo{howpublished}{\url{https://github.com/daanx/mimalloc-bench/}}.
\newblock


\bibitem[Mirosoft(2024)]%
        {mimalloc}
\bibfield{author}{\bibinfo{person}{Mirosoft}.} \bibinfo{year}{2024}\natexlab{}.
\newblock \bibinfo{title}{Mimalloc}.
\newblock \bibinfo{howpublished}{\url{https://github.com/microsoft/mimalloc/}}.
\newblock


\bibitem[Mittal(2017)]%
        {mittal2017survey}
\bibfield{author}{\bibinfo{person}{Sparsh Mittal}.}
  \bibinfo{year}{2017}\natexlab{}.
\newblock \showarticletitle{A survey of techniques for cache partitioning in
  multicore processors}.
\newblock \bibinfo{journal}{\emph{ACM Computing Surveys (CSUR)}}
  \bibinfo{volume}{50}, \bibinfo{number}{2} (\bibinfo{year}{2017}),
  \bibinfo{pages}{1--39}.
\newblock


\bibitem[Naffziger et~al\mbox{.}(2021)]%
        {naffziger2021pioneering}
\bibfield{author}{\bibinfo{person}{Samuel Naffziger}, \bibinfo{person}{Noah
  Beck}, \bibinfo{person}{Thomas Burd}, \bibinfo{person}{Kevin Lepak},
  \bibinfo{person}{Gabriel~H Loh}, \bibinfo{person}{Mahesh Subramony}, {and}
  \bibinfo{person}{Sean White}.} \bibinfo{year}{2021}\natexlab{}.
\newblock \showarticletitle{Pioneering chiplet technology and design for the
  amd epyc™ and ryzen™ processor families: Industrial product}. In
  \bibinfo{booktitle}{\emph{2021 ACM/IEEE 48th Annual International Symposium
  on Computer Architecture (ISCA)}}. IEEE, \bibinfo{pages}{57--70}.
\newblock


\bibitem[Nassif et~al\mbox{.}(2022)]%
        {nassif2022sapphire}
\bibfield{author}{\bibinfo{person}{Nevine Nassif}, \bibinfo{person}{Ashley~O
  Munch}, \bibinfo{person}{Carleton~L Molnar}, \bibinfo{person}{Gerald
  Pasdast}, \bibinfo{person}{Sitaraman~V Lyer}, \bibinfo{person}{Zibing Yang},
  \bibinfo{person}{Oscar Mendoza}, \bibinfo{person}{Mark Huddart},
  \bibinfo{person}{Srikrishnan Venkataraman}, \bibinfo{person}{Sireesha
  Kandula}, {et~al\mbox{.}}} \bibinfo{year}{2022}\natexlab{}.
\newblock \showarticletitle{Sapphire rapids: The next-generation intel xeon
  scalable processor}. In \bibinfo{booktitle}{\emph{2022 IEEE International
  Solid-State Circuits Conference (ISSCC)}}, Vol.~\bibinfo{volume}{65}. IEEE,
  \bibinfo{pages}{44--46}.
\newblock


\bibitem[Nvidia(2024)]%
        {grace}
\bibfield{author}{\bibinfo{person}{Nvidia}.} \bibinfo{year}{2024}\natexlab{}.
\newblock \bibinfo{title}{NVIDIA Grace Performance Tuning Guide}.
\newblock
  \bibinfo{howpublished}{\url{https://docs.nvidia.com/grace-performance-tuning-guide.pdf/}}.
\newblock


\bibitem[Oh et~al\mbox{.}(2021)]%
        {oh2021lctes}
\bibfield{author}{\bibinfo{person}{Deok-Jae Oh}, \bibinfo{person}{Yaebin Moon},
  \bibinfo{person}{Eojin Lee}, \bibinfo{person}{Tae~Jun Ham},
  \bibinfo{person}{Yongjun Park}, \bibinfo{person}{Jae~W. Lee}, {and}
  \bibinfo{person}{Jung~Ho Ahn}.} \bibinfo{year}{2021}\natexlab{}.
\newblock \showarticletitle{MaPHeA: A Lightweight Memory Hierarchy-Aware
  Profile-Guided Heap Allocation Framework}. In
  \bibinfo{booktitle}{\emph{Proceedings of the 22nd ACM SIGPLAN/SIGBED
  International Conference on Languages, Compilers, and Tools for Embedded
  Systems}} (Virtual, Canada) \emph{(\bibinfo{series}{LCTES 2021})}.
  \bibinfo{publisher}{Association for Computing Machinery},
  \bibinfo{address}{New York, NY, USA}, \bibinfo{pages}{24–36}.
\newblock
\showISBNx{9781450384728}
\href{https://doi.org/10.1145/3461648.3463844}{doi:\nolinkurl{10.1145/3461648.3463844}}


\bibitem[Pheatt(2008)]%
        {pheatt2008intel}
\bibfield{author}{\bibinfo{person}{Chuck Pheatt}.}
  \bibinfo{year}{2008}\natexlab{}.
\newblock \showarticletitle{Intel{\textregistered} threading building blocks}.
\newblock \bibinfo{journal}{\emph{Journal of Computing Sciences in Colleges}}
  \bibinfo{volume}{23}, \bibinfo{number}{4} (\bibinfo{year}{2008}),
  \bibinfo{pages}{298--298}.
\newblock


\bibitem[Rezaei and Kavi(2006)]%
        {rezaei2006intelligent}
\bibfield{author}{\bibinfo{person}{Mehran Rezaei} {and}
  \bibinfo{person}{Krishna~M Kavi}.} \bibinfo{year}{2006}\natexlab{}.
\newblock \showarticletitle{Intelligent memory manager: Reducing cache
  pollution due to memory management functions}.
\newblock \bibinfo{journal}{\emph{Journal of Systems Architecture}}
  \bibinfo{volume}{52}, \bibinfo{number}{1} (\bibinfo{year}{2006}),
  \bibinfo{pages}{41--55}.
\newblock


\bibitem[Sahraei et~al\mbox{.}(2023)]%
        {sahraei2023xfaas}
\bibfield{author}{\bibinfo{person}{Alireza Sahraei}, \bibinfo{person}{Soteris
  Demetriou}, \bibinfo{person}{Amirali Sobhgol}, \bibinfo{person}{Haoran
  Zhang}, \bibinfo{person}{Abhigna Nagaraja}, \bibinfo{person}{Neeraj Pathak},
  \bibinfo{person}{Girish Joshi}, \bibinfo{person}{Carla Souza},
  \bibinfo{person}{Bo Huang}, \bibinfo{person}{Wyatt Cook}, {et~al\mbox{.}}}
  \bibinfo{year}{2023}\natexlab{}.
\newblock \showarticletitle{Xfaas: Hyperscale and low cost serverless functions
  at meta}. In \bibinfo{booktitle}{\emph{Proceedings of the 29th Symposium on
  Operating Systems Principles}}. \bibinfo{pages}{231--246}.
\newblock


\bibitem[Sanchez and Kozyrakis(2011)]%
        {sanchez2011vantage}
\bibfield{author}{\bibinfo{person}{Daniel Sanchez} {and}
  \bibinfo{person}{Christos Kozyrakis}.} \bibinfo{year}{2011}\natexlab{}.
\newblock \showarticletitle{Vantage: Scalable and efficient fine-grain cache
  partitioning}. In \bibinfo{booktitle}{\emph{Proceedings of the 38th annual
  international symposium on Computer architecture}}. \bibinfo{pages}{57--68}.
\newblock


\bibitem[Schweizer et~al\mbox{.}(2015)]%
        {schweizer2015evaluating}
\bibfield{author}{\bibinfo{person}{Hermann Schweizer}, \bibinfo{person}{Maciej
  Besta}, {and} \bibinfo{person}{Torsten Hoefler}.}
  \bibinfo{year}{2015}\natexlab{}.
\newblock \showarticletitle{Evaluating the cost of atomic operations on modern
  architectures}. In \bibinfo{booktitle}{\emph{2015 International Conference on
  Parallel Architecture and Compilation (PACT)}}. IEEE,
  \bibinfo{pages}{445--456}.
\newblock


\bibitem[Sherwood et~al\mbox{.}(2002)]%
        {sherwood2002automatically}
\bibfield{author}{\bibinfo{person}{Timothy Sherwood}, \bibinfo{person}{Erez
  Perelman}, \bibinfo{person}{Greg Hamerly}, {and} \bibinfo{person}{Brad
  Calder}.} \bibinfo{year}{2002}\natexlab{}.
\newblock \showarticletitle{Automatically characterizing large scale program
  behavior}.
\newblock \bibinfo{journal}{\emph{ACM SIGPLAN Notices}} \bibinfo{volume}{37},
  \bibinfo{number}{10} (\bibinfo{year}{2002}), \bibinfo{pages}{45--57}.
\newblock


\bibitem[Sriraman and Dhanotia(2020)]%
        {sriraman2020accelerometer}
\bibfield{author}{\bibinfo{person}{Akshitha Sriraman} {and}
  \bibinfo{person}{Abhishek Dhanotia}.} \bibinfo{year}{2020}\natexlab{}.
\newblock \showarticletitle{Accelerometer: Understanding acceleration
  opportunities for data center overheads at hyperscale}. In
  \bibinfo{booktitle}{\emph{Proceedings of the Twenty-Fifth International
  Conference on Architectural Support for Programming Languages and Operating
  Systems}}. \bibinfo{pages}{733--750}.
\newblock


\bibitem[Stojkovic et~al\mbox{.}(2023)]%
        {stojkovic2023mumanycore}
\bibfield{author}{\bibinfo{person}{Jovan Stojkovic}, \bibinfo{person}{Chunao
  Liu}, \bibinfo{person}{Muhammad Shahbaz}, {and} \bibinfo{person}{Josep
  Torrellas}.} \bibinfo{year}{2023}\natexlab{}.
\newblock \showarticletitle{$\mu$Manycore: A Cloud-Native CPU for Tail at
  Scale}. In \bibinfo{booktitle}{\emph{Proceedings of the 50th Annual
  International Symposium on Computer Architecture}}. \bibinfo{pages}{1--15}.
\newblock


\bibitem[Tiwari et~al\mbox{.}(2010)]%
        {tiwari2010mmt}
\bibfield{author}{\bibinfo{person}{Devesh Tiwari}, \bibinfo{person}{Sanghoon
  Lee}, \bibinfo{person}{James Tuck}, {and} \bibinfo{person}{Yan Solihin}.}
  \bibinfo{year}{2010}\natexlab{}.
\newblock \showarticletitle{Mmt: Exploiting fine-grained parallelism in dynamic
  memory management}. In \bibinfo{booktitle}{\emph{2010 IEEE International
  Symposium on Parallel \& Distributed Processing (IPDPS)}}. IEEE,
  \bibinfo{pages}{1--12}.
\newblock


\bibitem[Wang et~al\mbox{.}(2016)]%
        {wang2016caf}
\bibfield{author}{\bibinfo{person}{Yipeng Wang}, \bibinfo{person}{Ren Wang},
  \bibinfo{person}{Andrew Herdrich}, \bibinfo{person}{James Tsai}, {and}
  \bibinfo{person}{Yan Solihin}.} \bibinfo{year}{2016}\natexlab{}.
\newblock \showarticletitle{CAF: Core to core communication acceleration
  framework}. In \bibinfo{booktitle}{\emph{Proceedings of the 2016
  International Conference on Parallel Architectures and Compilation}}.
  \bibinfo{pages}{351--362}.
\newblock


\bibitem[Wang et~al\mbox{.}(2024)]%
        {wang202430}
\bibfield{author}{\bibinfo{person}{Yipeng Wang}, \bibinfo{person}{Mengtian
  Yang}, \bibinfo{person}{Chieh-Pu Lo}, {and} \bibinfo{person}{Jaydeep~P
  Kulkarni}.} \bibinfo{year}{2024}\natexlab{}.
\newblock \showarticletitle{30.6 Vecim: A 289.13 GOPS/W RISC-V Vector
  Co-Processor with Compute-in-Memory Vector Register File for Efficient
  High-Performance Computing}. In \bibinfo{booktitle}{\emph{2024 IEEE
  International Solid-State Circuits Conference (ISSCC)}},
  Vol.~\bibinfo{volume}{67}. IEEE, \bibinfo{pages}{492--494}.
\newblock


\bibitem[Wang et~al\mbox{.}(2023)]%
        {Wang2023Memento}
\bibfield{author}{\bibinfo{person}{Ziqi Wang}, \bibinfo{person}{Kaiyang Zhao},
  \bibinfo{person}{Pei Li}, \bibinfo{person}{Andrew Jacob},
  \bibinfo{person}{Michael Kozuch}, \bibinfo{person}{Todd~C. Mowry}, {and}
  \bibinfo{person}{Dimitrios Skarlatos}.} \bibinfo{year}{2023}\natexlab{}.
\newblock \showarticletitle{Memento: Architectural Support for Ephemeral Memory
  Management in Serverless Environments}. In
  \bibinfo{booktitle}{\emph{MICRO-56: 56th Annual IEEE/ACM International
  Symposium on Microarchitecture}}.
\newblock
\href{https://doi.org/10.1145/3613424.3623795}{doi:\nolinkurl{10.1145/3613424.3623795}}


\bibitem[Wu et~al\mbox{.}(2021)]%
        {wu2021virtual}
\bibfield{author}{\bibinfo{person}{Qinzhe Wu}, \bibinfo{person}{Jonathan
  Beard}, \bibinfo{person}{Ashen Ekanayake}, \bibinfo{person}{Andreas
  Gerstlauer}, {and} \bibinfo{person}{Lizy~K John}.}
  \bibinfo{year}{2021}\natexlab{}.
\newblock \showarticletitle{Virtual-Link: A Scalable Multi-Producer
  Multi-Consumer Message Queue Architecture for Cross-Core Communication}. In
  \bibinfo{booktitle}{\emph{2021 IEEE International Parallel and Distributed
  Processing Symposium (IPDPS)}}. IEEE, \bibinfo{pages}{182--191}.
\newblock


\bibitem[Yang et~al\mbox{.}(2023)]%
        {yang2023numalloc}
\bibfield{author}{\bibinfo{person}{Hanmei Yang}, \bibinfo{person}{Xin Zhao},
  \bibinfo{person}{Jin Zhou}, \bibinfo{person}{Wei Wang},
  \bibinfo{person}{Sandip Kundu}, \bibinfo{person}{Bo Wu}, \bibinfo{person}{Hui
  Guan}, {and} \bibinfo{person}{Tongping Liu}.}
  \bibinfo{year}{2023}\natexlab{}.
\newblock \showarticletitle{Numalloc: A faster numa memory allocator}. In
  \bibinfo{booktitle}{\emph{Proceedings of the 2023 ACM SIGPLAN International
  Symposium on Memory Management}}. \bibinfo{pages}{97--110}.
\newblock


\bibitem[Zhou et~al\mbox{.}(2024)]%
        {zhou2024characterizing}
\bibfield{author}{\bibinfo{person}{Zhuangzhuang Zhou}, \bibinfo{person}{Vaibhav
  Gogte}, \bibinfo{person}{Nilay Vaish}, \bibinfo{person}{Chris Kennelly},
  \bibinfo{person}{Patrick Xia}, \bibinfo{person}{Svilen Kanev},
  \bibinfo{person}{Tipp Moseley}, \bibinfo{person}{Christina Delimitrou}, {and}
  \bibinfo{person}{Parthasarathy Ranganathan}.}
  \bibinfo{year}{2024}\natexlab{}.
\newblock \showarticletitle{Characterizing a Memory Allocator at Warehouse
  Scale}. In \bibinfo{booktitle}{\emph{ASPLOS}}.
\newblock


\end{thebibliography}
